\documentclass[aps,prd,epsfig,preprint,amsmath,amssymb,superscriptaddress,showkeys]{revtex4}
\usepackage{multirow}
\usepackage{graphicx,bm}
\usepackage[usenames]{color}
\usepackage{float}
\usepackage{caption}
\usepackage{subcaption}
\usepackage{slashed}
\begin{document}
\draft
\title{Study on the sensitivity of the Higgs boson couplings in photon-photon collision at CLIC and muon collider}

\author{S. Spor}
\email[]{serdar.spor@beun.edu.tr}
\affiliation{Department of Medical Imaging Techniques, Zonguldak B\"{u}lent Ecevit University, 67100, Zonguldak, T\"{u}rkiye.}

\begin{abstract}

In a model-independent way, we explore the potential of photon-induced interactions with the process $\gamma^* \gamma^* \rightarrow ZZ$ to investigate CP-conserving and CP-violating dimension-six operators of Higgs-gauge boson couplings using the Standard Model Effective Field Theory (SMEFT). The existence of anomalous $H\gamma\gamma$ and $HZZ$ couplings is discussed at 3 TeV Compact Linear Collider (CLIC) and 10 TeV Muon Collider (MuC) with integrated luminosities of 5 and 10 ab$^{-1}$, respectively. All signal and relevant background events are generated in MadGraph and passed through PYTHIA for parton showering and hadronization. The detector effects are evaluated using CLIC and MuC detector cards tuned in Delphes. We report the 95\% confidence level limits on the Wilson coefficients $\overline{c}_\gamma$, $\overline{c}_{HB}$, $\overline{c}_{HW}$, $\widetilde{c}_\gamma$, $\widetilde{c}_{HB}$, and $\widetilde{c}_{HW}$ and compare them with the experimental and phenomenological limits.

\end{abstract}


\maketitle

\section{Introduction} \label{Sec1}

The ATLAS \cite{Aad:2012les} and CMS \cite{Chatrchyan:2012les} collaborations at the Large Hadron Collider (LHC), along with the discovery of the 125 GeV Higgs boson, provided a gateway to the study of the dynamics of fundamental particles. This discovery clearly revealed the role of the Higgs mechanism in electroweak symmetry breaking (EWSB). Although experiments so far have shown that the Higgs boson is consistent with SM predictions, questions still remain about the true nature of the Higgs boson. As a result, subsequent studies have precisely measured the Higgs boson properties, that is, its association with SM particles and its CP nature. Considering the effective coupling of Higgs boson, with itself as well as with the gauge bosons and heavy fermions, making precise measurements of these couplings is crucial for determining the EWSB mechanism.

A more precise measurement of Higgs boson couplings paves the way for discovering the CP properties of the interactions between them. Although measurements \cite{Aad:2016saf,Aad:2020czq,Aad:2020jkn,Sirunyan:2020wxc,Tumasyan:2022ssx} show that the SM Higgs boson is a CP-even scalar with CP-conserving interactions and is consistent with the SM prediction, the presence of CP violation cannot be ruled out because of the matter-antimatter asymmetry observed in our Universe \cite{Steigman:1976ghb,Cohen:1987efv,Steigman:2008ujb}. There are two different types of CP violation in SM: the first is weak CP violation in the Cabibbo-Kobayashi-Maskawa (CKM) matrix \cite{Cabibbo:1963les,Kobayashi:1973hqw}, which defines the mixing of quark generations. The amount of CP violation from the CKM matrix in the SM is not sufficient to explain the observed matter-antimatter asymmetry of the Universe \cite{Riotto:1999jdx}. The second is the strong CP violation related to the topological charge in the Quantum Chromodynamics (QCD) vacuum \cite{Peccei:1977rve}. However, no CP violation was observed in any experiment that involved a strong interaction process.

CP violation is of great importance in understanding the matter-antimatter asymmetry or the baryon asymmetry of the Universe, which is one of the key problems that the SM struggles to explain, and is therefore one of the motivations that drives us to study beyond the SM. Electroweak baryogenesis (EWBG) is a mechanism that explains the matter-antimatter asymmetry of the Universe \cite{Kuzmin:1985yla}. It is particularly attractive as it is based only on electroweak scale physics and therefore can be tested experimentally. One of the three Sakharov conditions \cite{Sakharov:1967enp} required for a successful baryogenesis theory is C and CP violations. In the EWBG mechanism, CP violation plus sphaleron transition can produce baryon and lepton number violations during the electroweak phase transition, but the CP-violating ratio in the SM is too small to compensate for the amount of matter-antimatter asymmetry observed \cite{Feng:2023ecd}. Therefore, new additional sources of CP violation are needed to achieve electroweak baryogenesis, that is, to explain the matter-antimatter asymmetry in the Universe and naturally such sources exist in the extensions of the SM. Investigating Higgs boson interactions involving both CP-violating and CP-conserving couplings is notable for CP studies beyond the SM. However, the distinction between these two effects can be achieved when considering appropriate observables that allow us to probe the CP nature of Higgs couplings.

\section{Effective operators and Higgs boson interactions} \label{Sec2}

The Effective Field Theory (EFT) framework is used for the method of investigating Higgs couplings in a model-independent way. The EFT can only be valid up to a certain energy scale ($\Lambda$); this must be above the energy scale ($E$) directly accessible experimentally for the theory to be useful. That is, the EFT provides a good approximation when $\Lambda \gg E$. For interactions beyond the SM in the EFT framework, high-dimensional operators are used to parameterize the effects of unobserved states that are assumed to occur at energies larger than an effective scale defined by the $W$-boson mass $m_W$, in addition to the SM Lagrange.

The Strongly Interacting Light Higgs (SILH) basis consists of an effective theory of a light composite Higgs boson, which arises as the pseudo-Goldstone boson from a strongly-interacting sector and is responsible for the electroweak symmetry breaking \cite{Tosciri:2021yhb}. The effective Lagrangian, CP-conserving and CP-violating dimension-six operators of $H\gamma\gamma$ and $HZZ$ couplings in the SILH basis \cite{Giudice:2007ops} together with the dimension-four SM Lagrangian are defined as follows:

\begin{eqnarray}
\label{eq.1} 
{\cal L}_{\text{$H\gamma\gamma$, $HZZ$}}={\cal L}_{\text{SM}}+\sum_{i}{\overline{c}_i}{\cal O}_i={\cal L}_{\text{SM}}+{\cal L}_{\text{CPC}}+{\cal L}_{\text{CPV}}
\end{eqnarray}

{\raggedright where ${\cal O}_i$ are the dimension-six operators and ${\overline{c}_i}$ are the Wilson coefficients normalized with the new physics scale $\Lambda$ defined by the $W$-boson mass $m_W$. The second and third terms in Eq.~(\ref{eq.1}), which relate to the CP-conserving and CP-violating parts of the effective Lagrangian, are given below \cite{Alloul:2014hws}}

\begin{eqnarray}
\label{eq.2} 
\begin{split}
{\cal L}_{\text{CPC}}=&\frac{\overline{c}_H}{2\upsilon^2}\partial^\mu (\Phi^\dagger \Phi)\partial_\mu(\Phi^\dagger \Phi)+\frac{\overline{c}_T}{2\upsilon^2}(\Phi^\dagger \overset\leftrightarrow{D}^\mu \Phi)(\Phi^\dagger \overset\leftrightarrow{D}_\mu \Phi) \\
&+\frac{ig\overline{c}_W}{m_W^2}(\Phi^\dagger T_{2k}\overset\leftrightarrow{D}^\mu\Phi)D^\nu W_{\mu\nu}^k+\frac{ig^\prime \overline{c}_B}{2m_W^2}(\Phi^\dagger \overset\leftrightarrow{D}^\mu\Phi) \partial^\nu B_{\mu\nu} \\
&+\frac{2ig\overline{c}_{HW}}{m_W^2}(D^\mu \Phi^\dagger T_{2k}D^\nu\Phi)W_{\mu\nu}^k+\frac{ig^\prime \overline{c}_{HB}}{m_W^2}(D^\mu \Phi^\dagger D^\nu\Phi)B_{\mu\nu} \\
&+\frac{g^{\prime2}\overline{c}_{\gamma}}{m_W^2} \Phi^\dagger \Phi B_{\mu\nu} B^{\mu\nu}
\end{split}
\end{eqnarray}

{\raggedright and}

\begin{eqnarray}
\label{eq.3} 
{\cal L}_{\text{CPV}}=\frac{ig\widetilde{c}_{HW}}{m_W^2}D^\mu \Phi^\dagger T_{2k} {D}^\nu\Phi \widetilde{W}_{\mu\nu}^k+\frac{ig^\prime \widetilde{c}_{HB}}{m_W^2}D^\mu \Phi^\dagger {D}^\nu\Phi \widetilde{B}_{\mu\nu}+\frac{g^{\prime 2} \widetilde{c}_{\gamma}}{m_W^2}\Phi^\dagger \Phi B_{\mu\nu} \widetilde{B}^{\mu\nu}
\end{eqnarray}

{\raggedright where $\upsilon$ is the vacuum expectation value of the Higgs ﬁeld, $\widetilde{B}_{\mu\nu}=\frac{1}{2}\epsilon_{\mu\nu\rho\sigma}B^{\rho\sigma}$ and $\widetilde{W}_{\mu\nu}^k=\frac{1}{2}\epsilon_{\mu\nu\rho\sigma}W^{\rho\sigma k}$ are the dual field strength tensors defined by the field strength tensors ${B}_{\mu\nu}=\partial_\mu B_\nu - \partial_\nu B_\mu$ and ${W}_{\mu\nu}^k=\partial_\mu W_\nu^k - \partial_\nu W_\mu^k + g\epsilon_{ij}^k W_\mu^i W_\nu^j$ corresponding to $U(1)_Y$ and $SU(2)_L$ of the SM gauge groups, respectively, with gauge couplings $g^\prime$ and $g$. The generators of $SU(2)_L$ are given by $T_{2k}=\sigma_k/2$, where $\sigma_k$ is the Pauli matrices. $D^\mu$ is covariant derivative operator with $\Phi^\dagger \overset\leftrightarrow{D}_\mu\Phi=\Phi^\dagger(D_\mu \Phi)-(D_\mu \Phi^\dagger)\Phi$ and $\Phi$ is the Higgs doublet in SM.}

CP-conserving and CP-violating dimension-six operators in the SILH basis given by Eqs.~(\ref{eq.2}-\ref{eq.3}) define in terms of the mass-eigenstates after the electroweak symmetry breaking. The Lagrangian with anomalous $H\gamma\gamma$ and $HZZ$ couplings in the unitarity gauge and in the mass basis is given by \cite{Alloul:2014hws}    

\begin{eqnarray}
\label{eq.4} 
\begin{split}
{\cal L}=&-\frac{1}{4}g_{h\gamma\gamma}F_{\mu\nu}F^{\mu\nu}h-\frac{1}{4}\widetilde{g}_{h\gamma\gamma}F_{\mu\nu}\widetilde{F}^{\mu\nu}h-\frac{1}{4}g_{hzz}^{(1)}Z_{\mu\nu}Z^{\mu\nu}h-g_{hzz}^{(2)}Z_{\nu}\partial_\mu Z^{\mu\nu}h \\
&+\frac{1}{2}g_{hzz}^{(3)}Z_{\mu}Z^{\mu}h-\frac{1}{4}\widetilde{g}_{hzz}Z_{\mu\nu}\widetilde{Z}^{\mu\nu}h
\end{split}
\end{eqnarray}     

{\raggedright where $h$ is the Higgs-boson field and the ﬁeld strength tensors of photon and $Z$-boson are introduced with $F_{\mu\nu}$ and $Z_{\mu\nu}$. $\widetilde{F}^{\mu\nu}=\frac{1}{2}\epsilon^{\mu\nu\rho\sigma}F_{\rho\sigma}$ and $\widetilde{Z}^{\mu\nu}=\frac{1}{2}\epsilon^{\mu\nu\rho\sigma}Z_{\rho\sigma}$ are the dual field strength tensors. The relations between the Lagrangian parameters in the gauge basis given in Eqs.~(\ref{eq.2}-\ref{eq.3}) and in the mass basis given in Eq.~(\ref{eq.4}), which express the interactions of a Higgs boson with a vector boson pair, are presented below for CP-violating couplings}

\begin{eqnarray}
\label{eq.5} 
\widetilde{g}_{h\gamma\gamma}=-\frac{8g\widetilde{c}_\gamma s_W^2}{m_W}
\end{eqnarray}
\begin{eqnarray}
\label{eq.6} 
\widetilde{g}_{hzz}=\frac{2g}{c_W^2 m_W}\left[\widetilde{c}_{HB}s_W^2-4\widetilde{c}_{\gamma}s_W^4+c_W^2\widetilde{c}_{HW}\right]
\end{eqnarray}

{\raggedright and for CP-conserving couplings}

\begin{eqnarray}
\label{eq.7} 
g_{h\gamma\gamma}=a_H-\frac{8g\overline{c}_\gamma s_W^2}{m_W}
\end{eqnarray}
\begin{eqnarray}
\label{eq.8} 
g_{hzz}^{(1)}=\frac{2g}{c_W^2 m_W}\left[\overline{c}_{HB}s_W^2-4\overline{c}_{\gamma}s_W^4+c_W^2 \overline{c}_{HW}\right]
\end{eqnarray}
\begin{eqnarray}
\label{eq.9} 
g_{hzz}^{(2)}=\frac{g}{c_W^2 m_W}\left[(\overline{c}_{HW}+\overline{c}_{W})c_W^2+(\overline{c}_B+\overline{c}_{HB})s_W^2\right]
\end{eqnarray}
\begin{eqnarray}
\label{eq.10} 
g_{hzz}^{(3)}=\frac{gm_W}{c_W^2}\left[1-\frac{1}{2}\overline{c}_{H}-2\overline{c}_{T}+8\overline{c}_\gamma \frac{s_W^4}{c_W^2}\right]
\end{eqnarray}

{\raggedright where $s_W=\text{sin}\theta_W$ and $c_W=\text{cos}\theta_W$ with $\theta_W$ being the weak mixing angle. $a_H$ indicates the SM contributions for the Higgs boson to two photons vertex. There are eight Wilson coefficients in total for $H\gamma\gamma$ and $HZZ$ couplings. The coefficients $\overline{c}_\gamma$, $\overline{c}_{HB}$, $\overline{c}_{HW}$, $\overline{c}_B$, $\overline{c}_W$, $\overline{c}_H$, $\overline{c}_T$ relate to CP-conserving couplings, while coefficients $\widetilde{c}_\gamma$, $\widetilde{c}_{HB}$, $\widetilde{c}_{HW}$ relate to CP-violating couplings. According to Refs.~\cite{Ellis:2015pec,Khanpour:2017ubw}, the setting of $\overline{c}_B=-\overline{c}_W$ is allowed by restricting on the combination $\overline{c}_W+\overline{c}_B$. The $Z$-boson pair production with the process $\gamma^* \gamma^* \rightarrow ZZ$ is sensitive to Higgs-gauge boson couplings in Eqs.~(\ref{eq.5}-\ref{eq.10}).}

In this study, analyses of the effects of dimension-six operators in Higgs-gauge boson couplings are performed into {\sc MadGraph5}$\_$aMC@NLO \cite{Alwall:2014cvc} based on Monte Carlo simulations using FeynRules \cite{Alloul:2014tfc} and the UFO framework. The SILH Lagrangian in Eq.~(\ref{eq.1}) is contained in the Higgs Effective Lagrangian (HEL) model file of the FeynRules and the HEL model includes 39 Wilson coefficients for all interactions.

We investigate the potential of anomalous Higgs-gauge boson couplings on the process $\gamma^* \gamma^* \rightarrow ZZ \rightarrow \ell \ell \nu \nu$ at the Compact Linear Collider (CLIC) and the Muon Collider (MuC). We focus on the sensitivity study of $\overline{c}_\gamma$, $\overline{c}_{HB}$, $\overline{c}_{HW}$, $\widetilde{c}_\gamma$, $\widetilde{c}_{HB}$ and $\widetilde{c}_{HW}$ coefficients at anomalous $H\gamma\gamma$ and $HZZ$ vertices resulting from $Z$-boson pair production through a photon-induced process and the examination of the CP properties of these couplings.

The following parameters are assumed for these two colliders: $\sqrt{s}=3$ TeV with ${\cal L}_{\text{int}}=5$ ab$^{-1}$ for the CLIC \cite{Robson:2018tgb} and $\sqrt{s}=10$ TeV with ${\cal L}_{\text{int}}=10$ ab$^{-1}$ for the MuC \cite{Aime:2022tqz}. The fact that lepton colliders have a cleaner environment than hadron colliders with many backgrounds provides obvious benefits in maximizing the physics potential for precision Higgs studies. Recently, numerous studies have been performed to investigate the measurement of Higgs boson couplings at the CLIC \cite{Abramowicz:2017wxz,Ellis:2017pmw,Denizli:2018rca,Sahin:2019unz,Karadeniz:2020yvz,Roloff:2020efg} and the MuC \cite{Chiesa:2020yhn,Han:2021pas,Chen:2021pln,Han:2021pmq,Chen:2022ygc,Forslund:2022wzc,Blas:2022tzs}.

In our study, there are $\gamma^* \gamma^*$ collisions in the processes $e^+e^-\rightarrow e^+\gamma^*\gamma^*e^-\rightarrow e^+ZZe^-$ and $\mu^+\mu^-\rightarrow\mu^+\gamma^*\gamma^*\mu^- \rightarrow\mu^+ZZ\mu^-$. The $\gamma^*$ in $\gamma^*\gamma^*$ collisions are quasi-real photons, which can be defined by the Equivalent Photon Approximation (EPA) \cite{Budnev:1975kyp,Piotrzkowski:2001jks}. According to the EPA, quasi-real photons with very low virtuality emitted from incoming $e^+(e^-)$ or $\mu^+(\mu^-)$ beams at $e^+e^-$ or $\mu^+\mu^-$ colliders scatter at very small angles from the beam tube. These EPA processes have been experimentally observed in LEP, Tevatron and LHC \cite{Abulencia:2007gas,Aaltonen:2009unz,Aaltonen:2009yaz,Chatrchyan:2012tcz,Chatrchyan:2012gvz,Abazov:2013qsa,Chatrchyan:2013tlk}. In the EPA, the spectrum of photon emitted from electron (muon) is given by:

\begin{eqnarray}
\label{eq.11}
\begin{split}
f_{\gamma^{*}}(x)=&\, \frac{\alpha}{\pi E_{e(\mu)}}\Bigg\{\left[\frac{1-x+x^{2}/2}{x}\right]\text{log}\left(\frac{Q_{\text{max}}^{2}}{Q_{\text{min}}^{2}}\right)-\frac{m_{e(\mu)}^{2}x}{Q_{\text{min}}^{2}}\left(1-\frac{Q_{\text{min}}^{2}}{Q_{\text{max}}^{2}}\right)
\\
&-\frac{1}{x}\left[1-\frac{x}{2}\right]^{2}\text{log}\left(\frac{x^{2}E_{e(\mu)}^{2}+Q_{\text{max}}^{2}}{x^{2}E_{e(\mu)}^{2}+Q_{\text{min}}^{2}}\right)\Bigg\}\,,
\end{split}
\end{eqnarray}

{\raggedright where $\alpha$ is the ﬁne-structure constant, $x = E_{\gamma^*}/E_{e(\mu)}$, $m_{e(\mu)}$ is the mass of the scattering particles, $Q^2_{\text{min}}=m^2_{e(\mu)}x^2/(1-x)$ and $Q^2_{\text{max}}$ are the minimum and maximum virtuality of photon, $E_{\gamma^*}$ and $E_{e(\mu)}$ are the energies of photon and scattered electron (muon).}

\begin{figure}[H]
\centering
\includegraphics[scale=0.66]{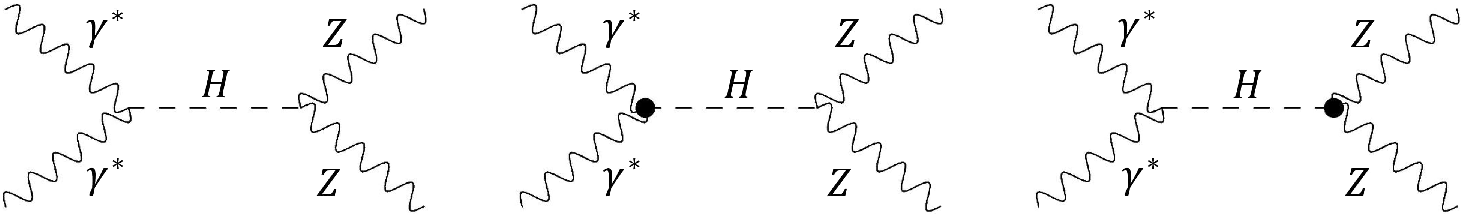}
\caption{The Feynman diagrams of the process $\gamma^* \gamma^* \rightarrow ZZ$. 
\label{fig1}}
\end{figure}

The Feynman diagrams of the process $\gamma^* \gamma^* \rightarrow ZZ$ are shown in Fig.~\ref{fig1}. The diagram on the left presents only the SM background process, while the diagrams on the middle and right represent the signal processes including the anomalous $H\gamma\gamma$ and $HZZ$ vertices, respectively.

\begin{figure}[H]
\centering
\begin{subfigure}{0.48\linewidth}
\includegraphics[width=\linewidth]{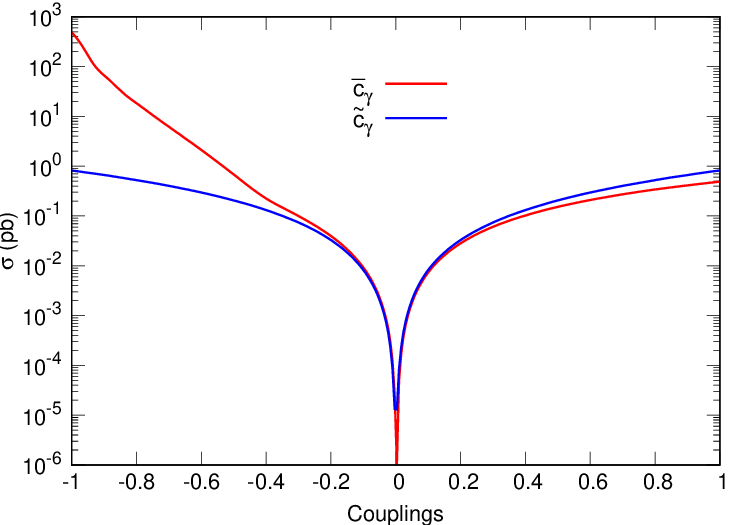}
\caption{}
\label{fig2:a}
\end{subfigure}\hfill
\begin{subfigure}{0.48\linewidth}
\includegraphics[width=\linewidth]{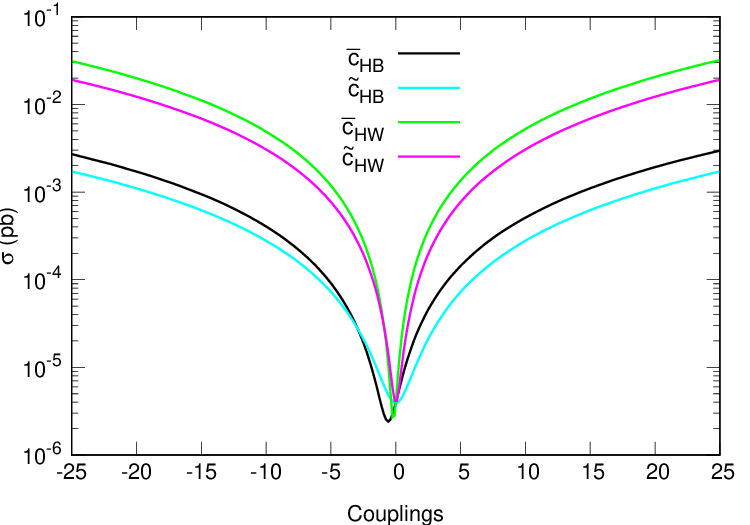}
\caption{}
\label{fig2:b}
\end{subfigure}\hfill
\begin{subfigure}{0.48\linewidth}
\includegraphics[width=\linewidth]{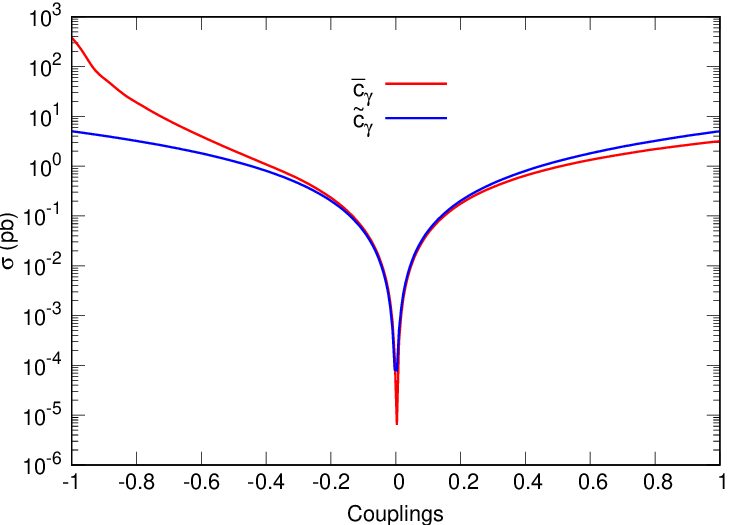}
\caption{}
\label{fig2:c}
\end{subfigure}\hfill
\begin{subfigure}{0.48\linewidth}
\includegraphics[width=\linewidth]{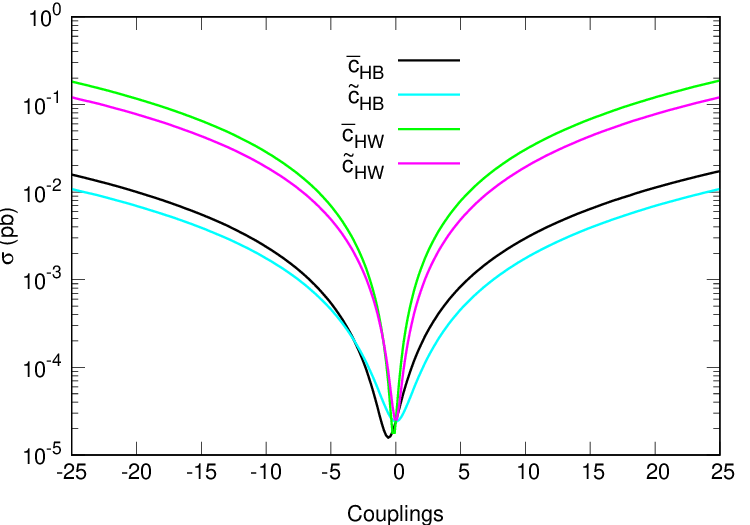}
\caption{}
\label{fig2:d}
\end{subfigure}\hfill
\caption{The total cross-section as a function of $\overline{c}_\gamma$, $\widetilde{c}_\gamma$ coefficients (left) and $\overline{c}_{HB}$, $\widetilde{c}_{HB}$, $\overline{c}_{HW}$, $\widetilde{c}_{HW}$ coefficients (right) at the CLIC (top row) and the MuC (bottom row).}
\label{fig2}
\end{figure}

The total cross-sections of the process $\gamma^* \gamma^* \rightarrow ZZ \rightarrow \ell \ell \nu \nu$ as a function of $\overline{c}_\gamma$, $\overline{c}_{HB}$, $\overline{c}_{HW}$, $\widetilde{c}_\gamma$, $\widetilde{c}_{HB}$ and $\widetilde{c}_{HW}$ coefficients at the CLIC and the MuC are shown in Fig.~\ref{fig2}. The total cross-sections in Fig.~\ref{fig2} are calculated by applying the transverse momentum $p_T^{\ell^1,\ell^2} > 10$ GeV and the pseudo-rapidity $|\eta^{\ell^1,\ell^2}| < 2.5$ for the charged leptons on the final state particles. The calculation method in these figures is that the coefficient under consideration is variable each time, while the other coefficients are fixed to zero. The cross-sections of $3.94\times 10^{-6}$ pb at CLIC and $2.41\times 10^{-5}$ pb at MuC resulting from the SM contributions from the Feynman diagram on the left in Fig.~\ref{fig1} are numerically calculated at the points $\overline{c}_\gamma=\widetilde{c}_\gamma=\overline{c}_{HB}=\overline{c}_{HW}=\widetilde{c}_{HB}=\widetilde{c}_{HW}=0$. It is seen that the effect of $\overline{c}_\gamma$ and $\widetilde{c}_\gamma$ coefficients is greater than that of $\overline{c}_{HB}$, $\overline{c}_{HW}$, $\widetilde{c}_{HB}$ and $\widetilde{c}_{HW}$ coefficients, according to the total cross-sections as a function of the coefficients. Therefore, in the analyses, the values of $\overline{c}_{HB}$, $\overline{c}_{HW}$, $\widetilde{c}_{HB}$ and $\widetilde{c}_{HW}$ coefficients will be considered higher than the values of $\overline{c}_\gamma$ and $\widetilde{c}_\gamma$ coefficients.

\section{Signal and Background Analysis} \label{Sec3}

We present details of the simulation and the cut-based analysis to obtain the limits on anomalous Higgs-gauge boson couplings at the $H\gamma\gamma$ and $HZZ$ vertices via the photon-induced process at the CLIC and the MuC. The process $\gamma^* \gamma^* \rightarrow ZZ$ with non-zero $\overline{c}_\gamma$, $\overline{c}_{HB}$, $\overline{c}_{HW}$, $\widetilde{c}_\gamma$, $\widetilde{c}_{HB}$, $\widetilde{c}_{HW}$ coefficients is considered as signal including the SM contribution as well as interference between effective couplings and SM contributions $(S+B1)$. We consider twelve relevant backgrounds $(B1-B12)$ at $e^+e^-$ or $\mu^+\mu^-$ colliders; backgrounds $B1-B6$ are in photon-induced collisions, while backgrounds $B7-B12$ are not in photon-induced collisions. These are defined as follows: the SM contribution $(B1)$ with the same final state of the considered signal process ($\gamma^* \gamma^* \rightarrow ZZ \rightarrow \ell \ell \nu \nu$) and the other SM background contributions from $\gamma^* \gamma^* \rightarrow W^-W^+ \rightarrow \ell^- \bar{\nu} \ell^+ \nu$ process with $W$-boson pair production $(B2)$, $\gamma^* \gamma^* \rightarrow W^-W^+Z \rightarrow \ell^- \bar{\nu} \ell^+ \nu \nu \bar{\nu}$ process with $WWZ$ production $(B3)$, $\gamma^* \gamma^* \rightarrow \nu \bar{\nu} Z \rightarrow \nu \bar{\nu} \ell^- \ell^+$ process $(B4)$, $\gamma^* \gamma^* \rightarrow \tau^+\tau^- \rightarrow \ell^+ \nu \bar{\nu}_\tau \ell^- \bar{\nu} \nu_\tau$ process with tau-pair production $(B5)$, $\gamma^* \gamma^* \rightarrow t\bar{t} \rightarrow \ell^+ \nu b \ell^- \bar{\nu} \bar{b}$ process with top-quark pair production $(B6)$, $e^+e^-/\mu^+\mu^- \rightarrow ZZ \rightarrow \ell \ell \nu \nu$ process $(B7)$, $e^+e^-/\mu^+\mu^- \rightarrow W^-W^+ \rightarrow \ell^- \bar{\nu} \ell^+ \nu$ process $(B8)$, $e^+e^-/\mu^+\mu^- \rightarrow W^-W^+Z \rightarrow \ell^- \bar{\nu} \ell^+ \nu \nu \bar{\nu}$ process $(B9)$, $e^+e^-/\mu^+\mu^- \rightarrow \nu \bar{\nu} Z \rightarrow \nu \bar{\nu} \ell^- \ell^+$ process $(B10)$, $e^+e^-/\mu^+\mu^- \rightarrow \tau^+\tau^- \rightarrow \ell^+ \nu \bar{\nu}_\tau \ell^- \bar{\nu} \nu_\tau$ process $(B11)$ and $e^+e^-/\mu^+\mu^- \rightarrow t\bar{t} \rightarrow \ell^+ \nu b \ell^- \bar{\nu} \bar{b}$ process $(B12)$. Since tau leptons, which have a lifetime of $3\times10^{-13}$ s, decay before being detected, we simply refer to charged leptons as electrons $(e^\pm)$ and muons $(\mu^\pm)$ in the signal and background analyses. 500k events are generated for each coupling value of the signal processes and each of the background processes in {\sc MadGraph5}$\_$aMC@NLO. These events are passed through the Pythia 8.3 \cite{Bierlich:2022uzx} including initial and ﬁnal parton shower, the fragmentation and decay. The detector responses are simulated using Delphes 3.5.0 package \cite{Favereau:2014wfb} with the CLIC and the MuC configuration cards and all events are analyzed with ROOT 6 \cite{Brun:1997gqa}.

Preselection and kinematic cuts are applied separately to the $\ell\ell\nu\nu$ channel in both the signal and background processes to distinguish signal from relevant backgrounds. In order to determine the cuts, it is necessary to identify the region where the signal has a distinctive signature relative to the backgrounds. Therefore, we consider various kinematic distributions for the signal and interfering background depending on the cuts. The presence of an opposite-sign same flavor dilepton in the decay of the $Z$-boson is determined as the preselection preference in the analysis and is labeled with Cut-0.

\begin{figure}[H]
\centering
\begin{subfigure}{0.48\linewidth}
\includegraphics[width=\linewidth]{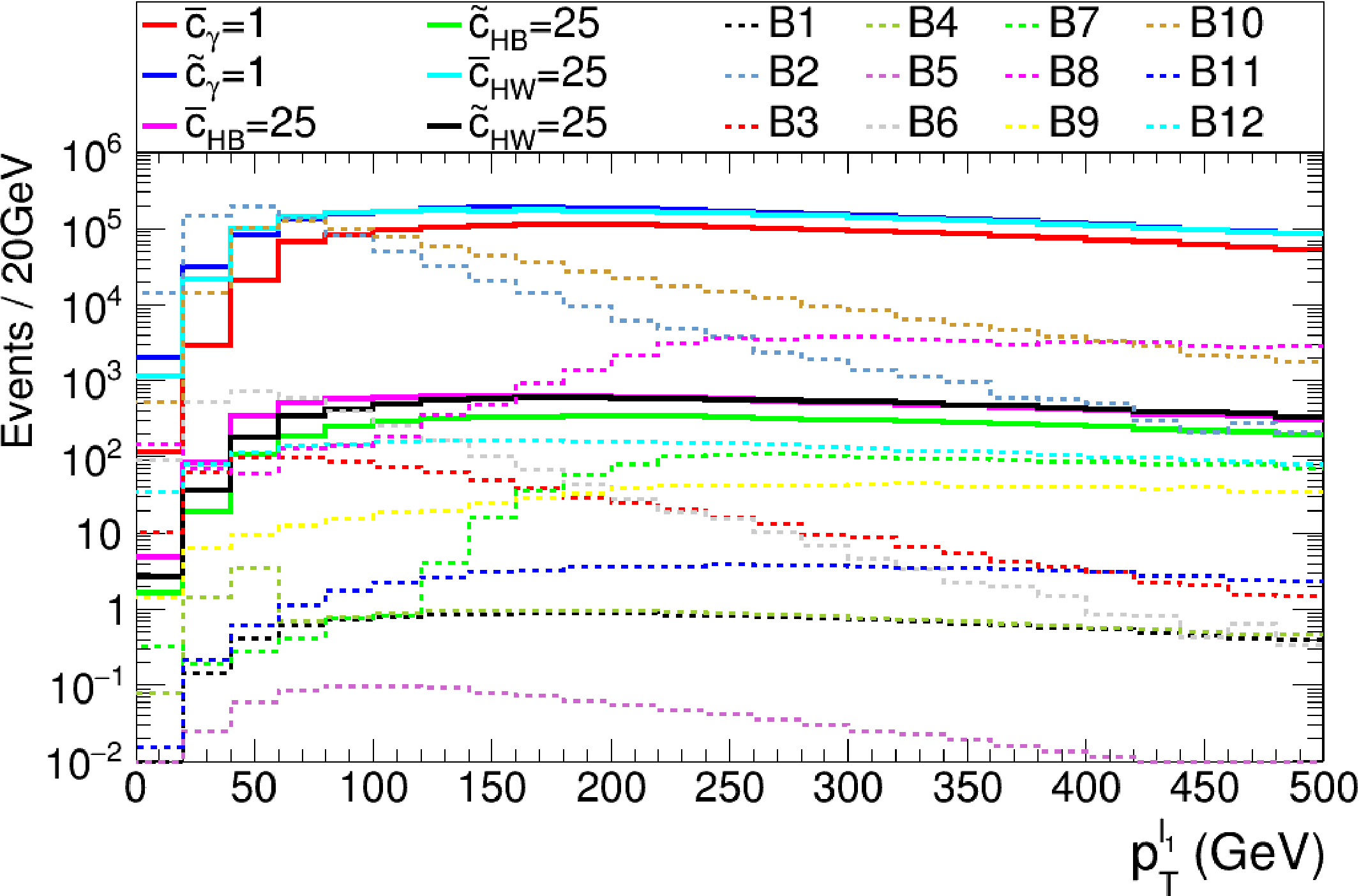}
\caption{}
\label{fig3:a}
\end{subfigure}\hfill
\begin{subfigure}{0.48\linewidth}
\includegraphics[width=\linewidth]{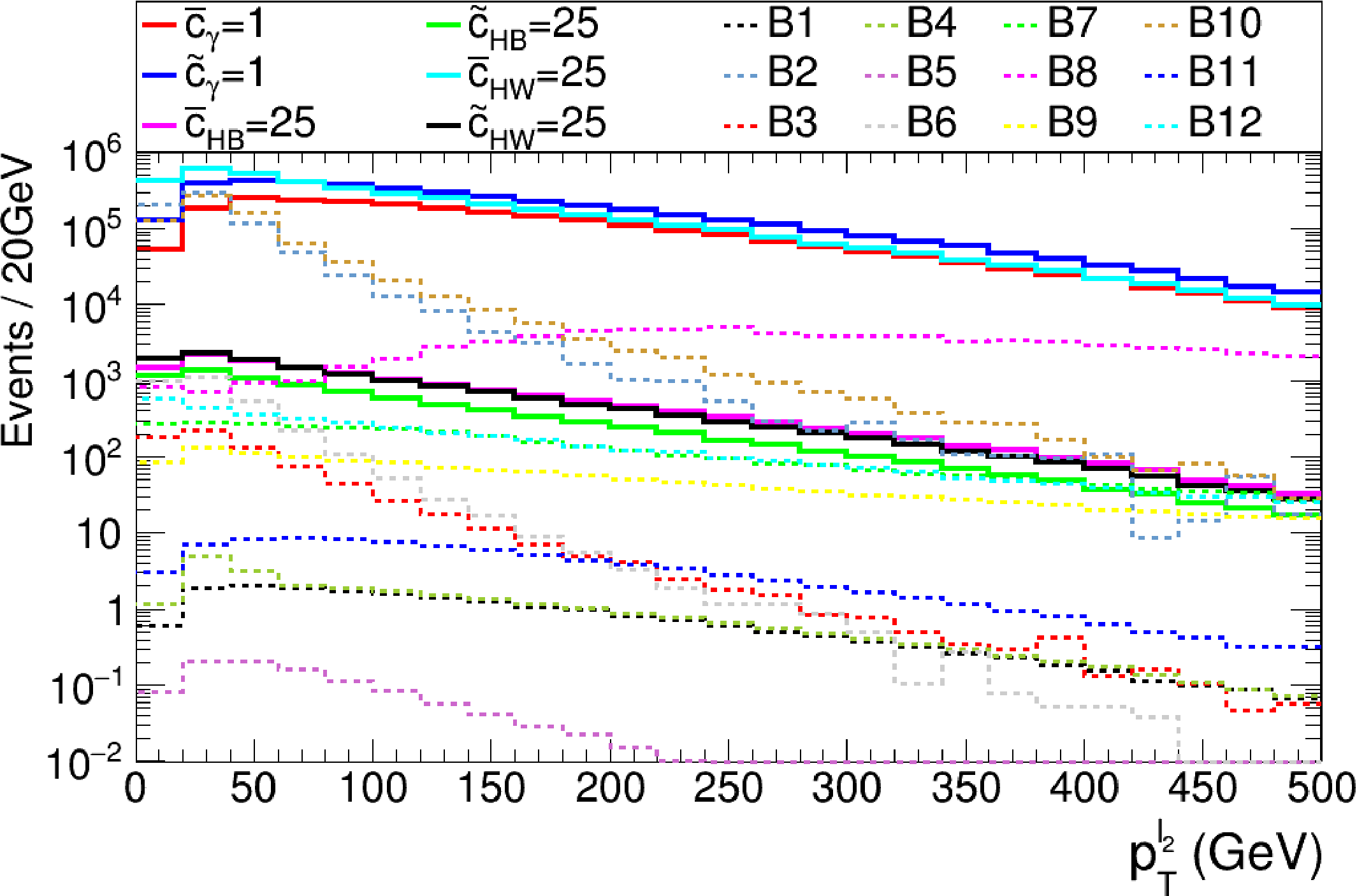}
\caption{}
\label{fig3:b}
\end{subfigure}\hfill
\begin{subfigure}{0.48\linewidth}
\includegraphics[width=\linewidth]{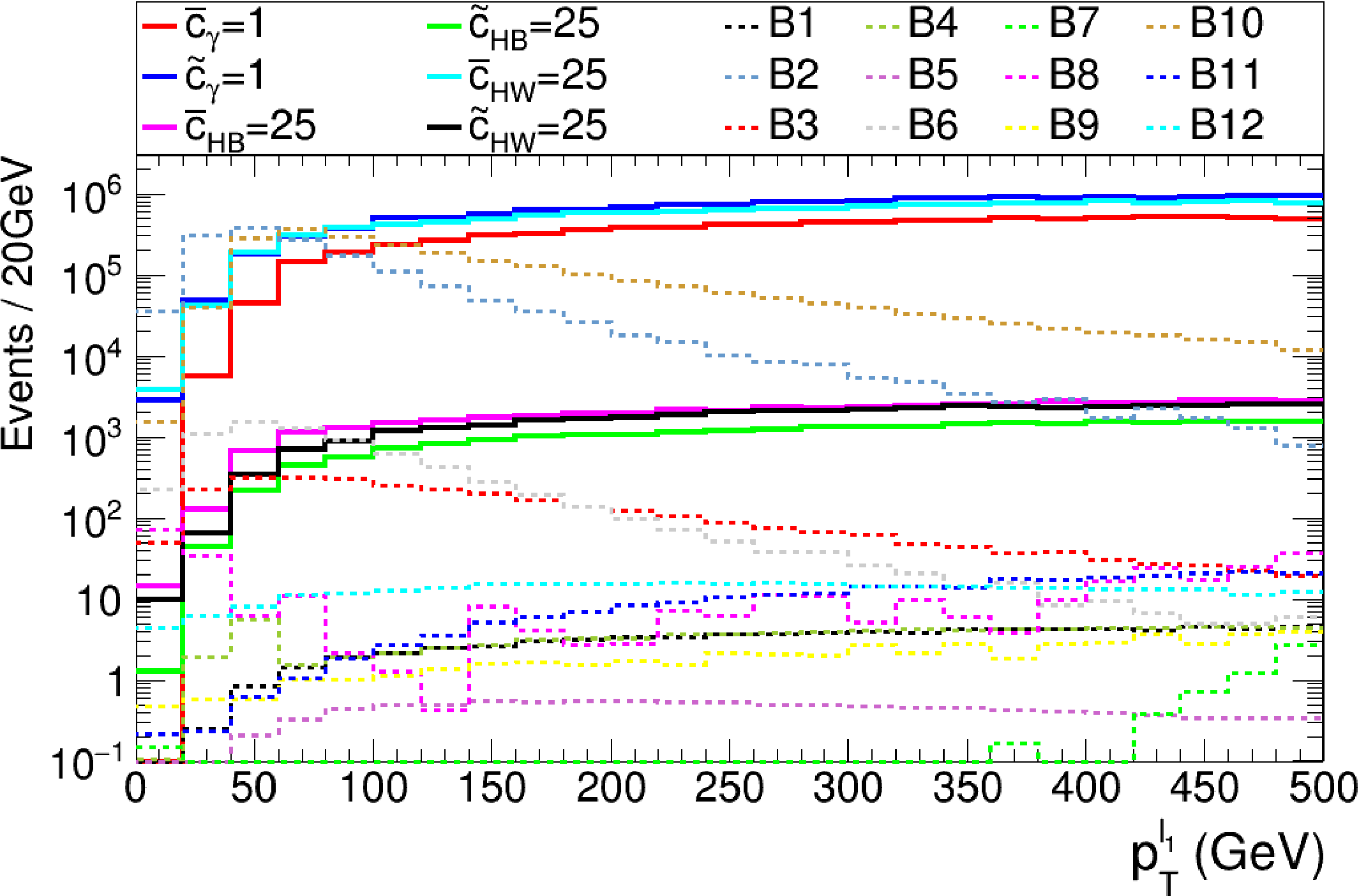}
\caption{}
\label{fig3:c}
\end{subfigure}\hfill
\begin{subfigure}{0.48\linewidth}
\includegraphics[width=\linewidth]{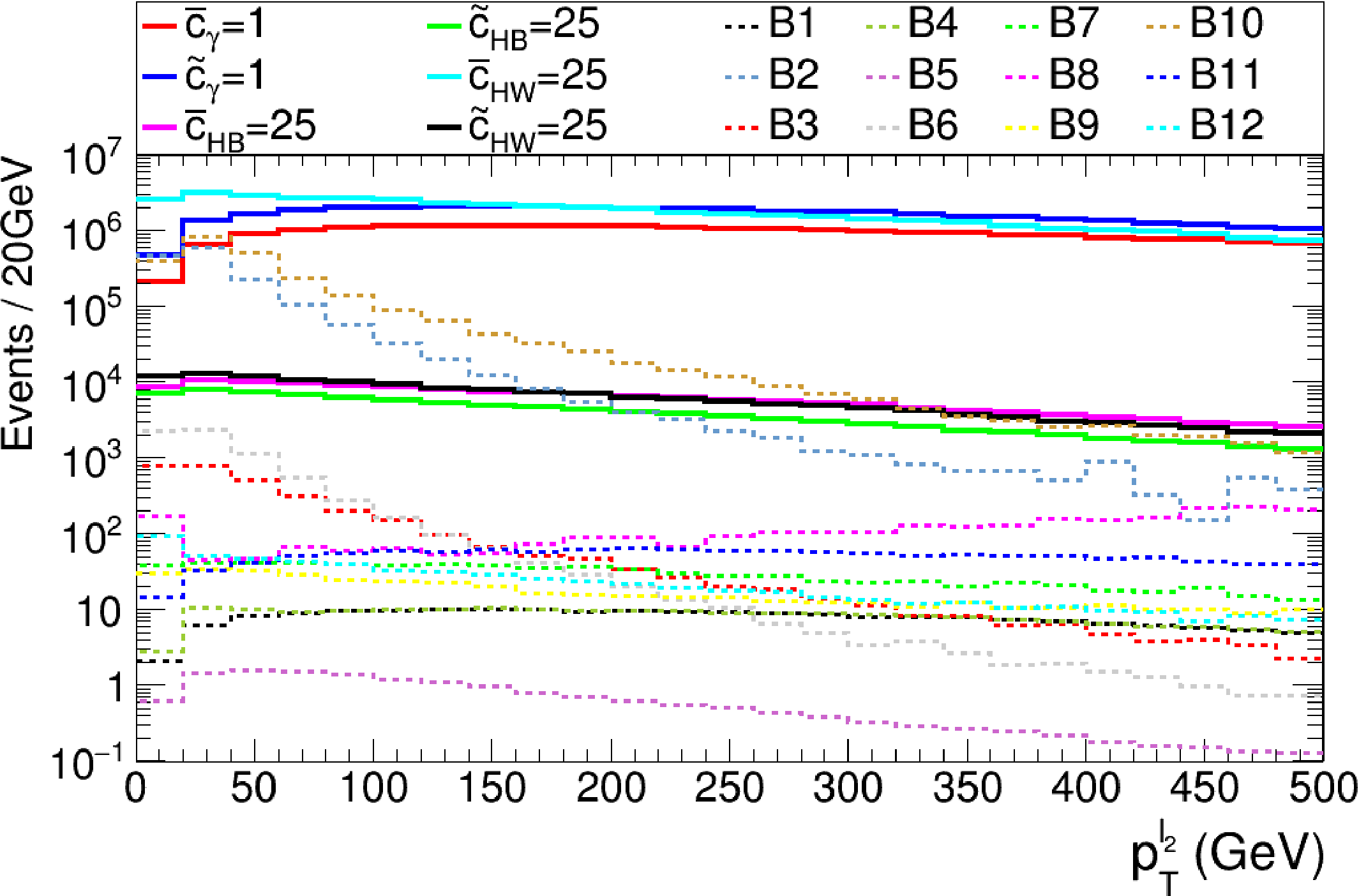}
\caption{}
\label{fig3:d}
\end{subfigure}\hfill

\caption{The transverse momentum distributions of the leading charged lepton (left) and the sub-leading charged lepton (right) for signal with $\overline{c}_\gamma=1$, $\overline{c}_{HB}=25$, $\overline{c}_{HW}=25$, $\widetilde{c}_\gamma=1$, $\widetilde{c}_{HB}=25$, $\widetilde{c}_{HW}=25$ and relevant background
processes at the CLIC (top row) and the MuC (bottom row).}
\label{fig3}
\end{figure}

\begin{figure}[H]
\centering
\begin{subfigure}{0.48\linewidth}
\includegraphics[width=\linewidth]{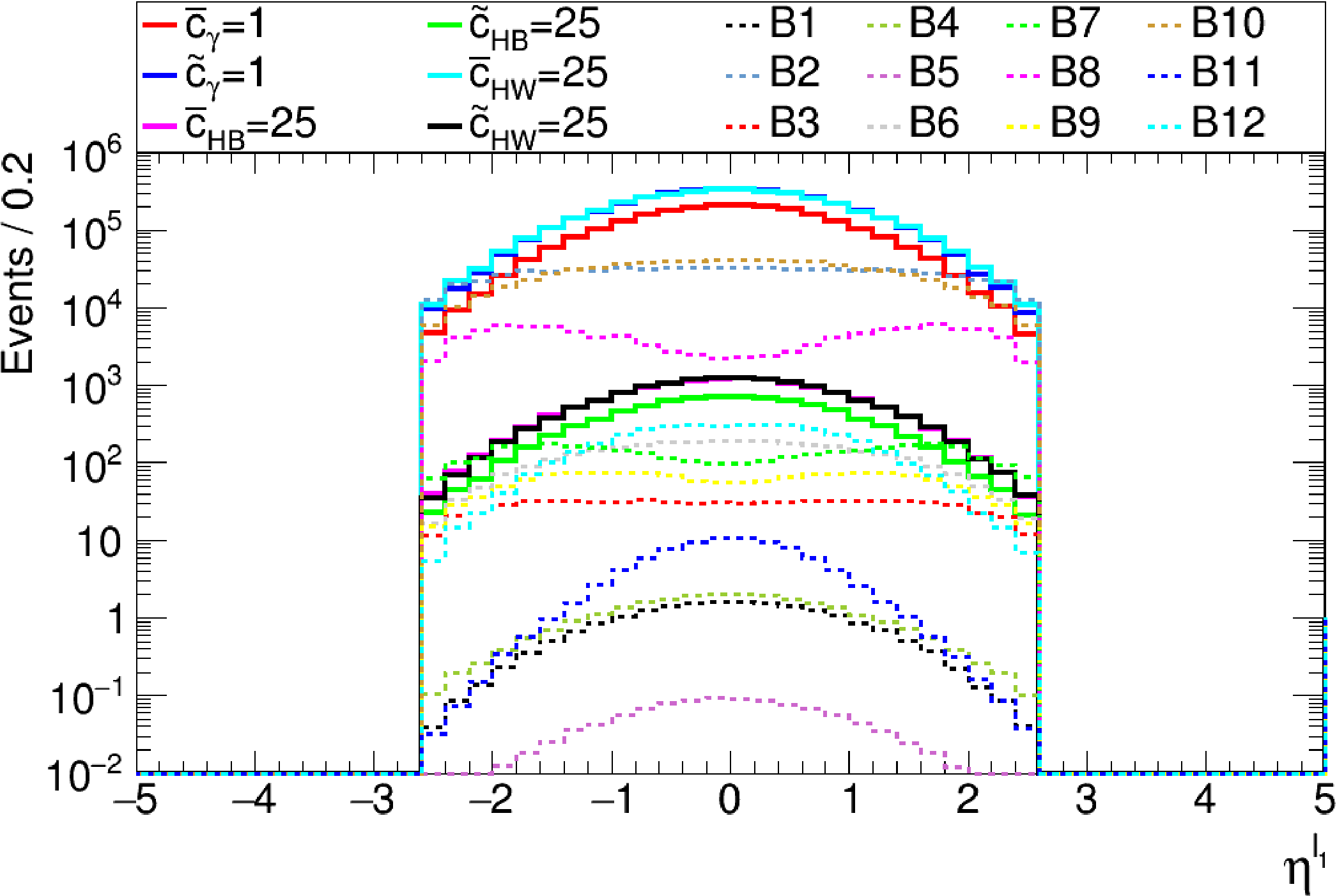}
\caption{}
\label{fig4:a}
\end{subfigure}\hfill
\begin{subfigure}{0.48\linewidth}
\includegraphics[width=\linewidth]{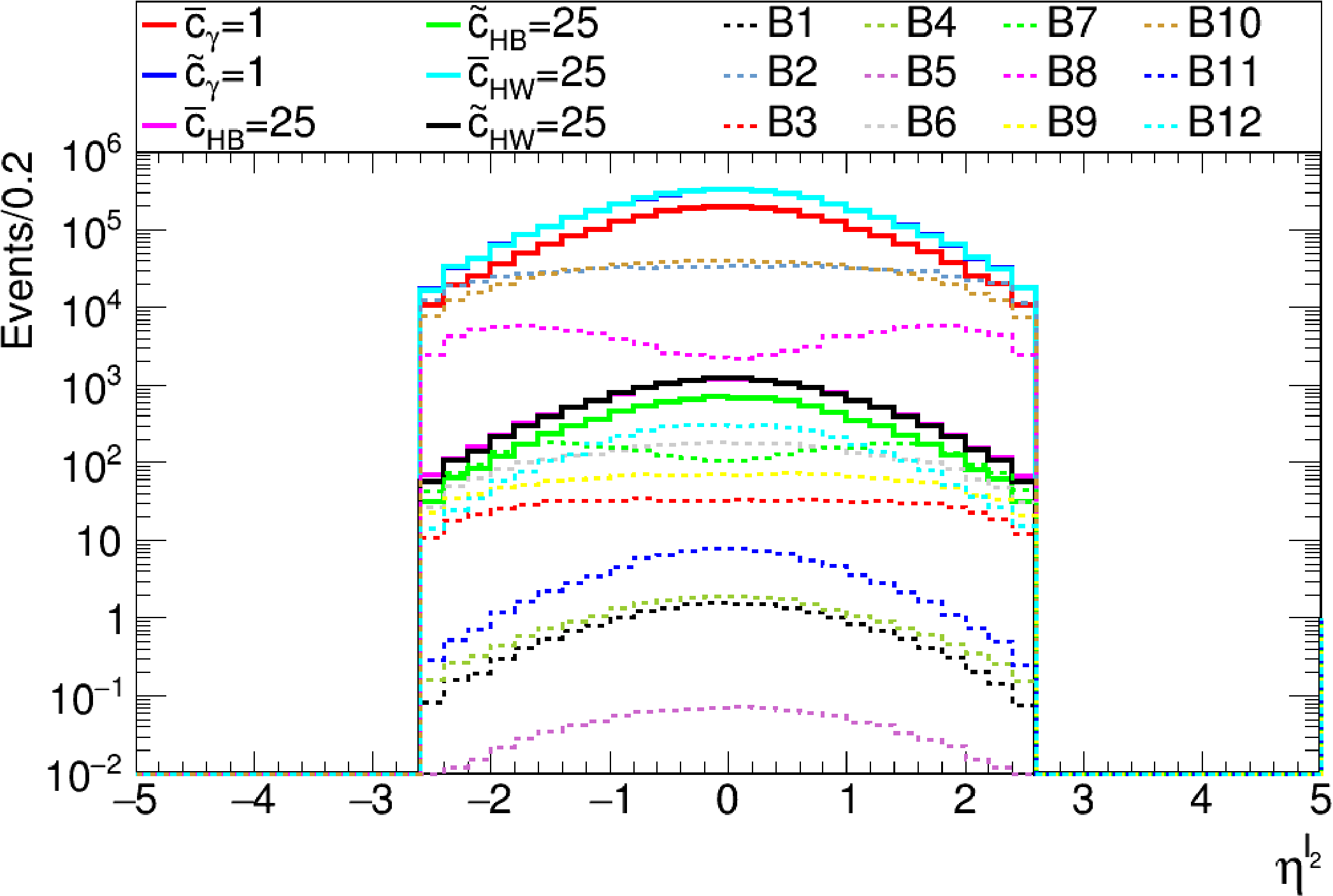}
\caption{}
\label{fig4:b}
\end{subfigure}\hfill
\begin{subfigure}{0.48\linewidth}
\includegraphics[width=\linewidth]{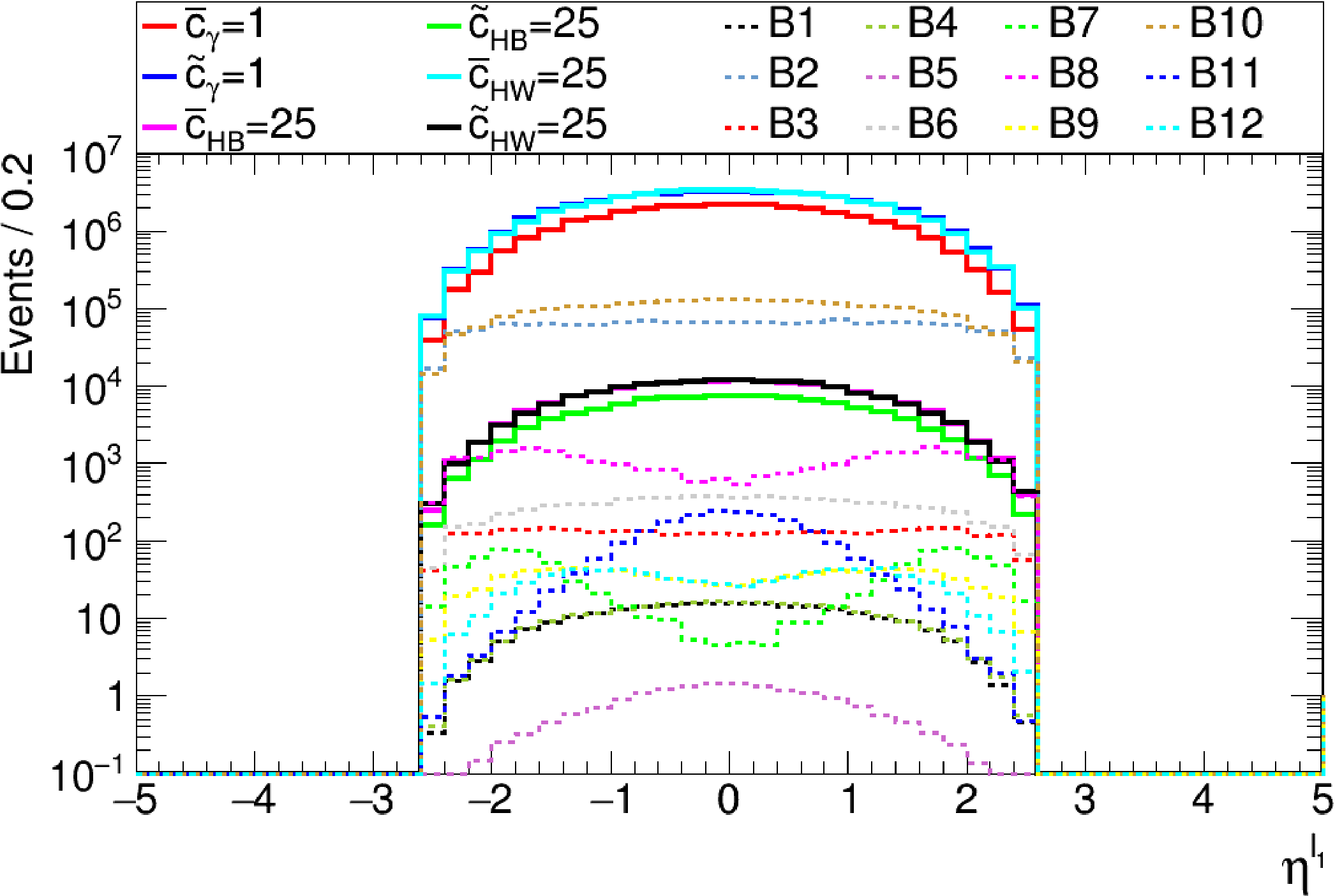}
\caption{}
\label{fig4:c}
\end{subfigure}\hfill
\begin{subfigure}{0.48\linewidth}
\includegraphics[width=\linewidth]{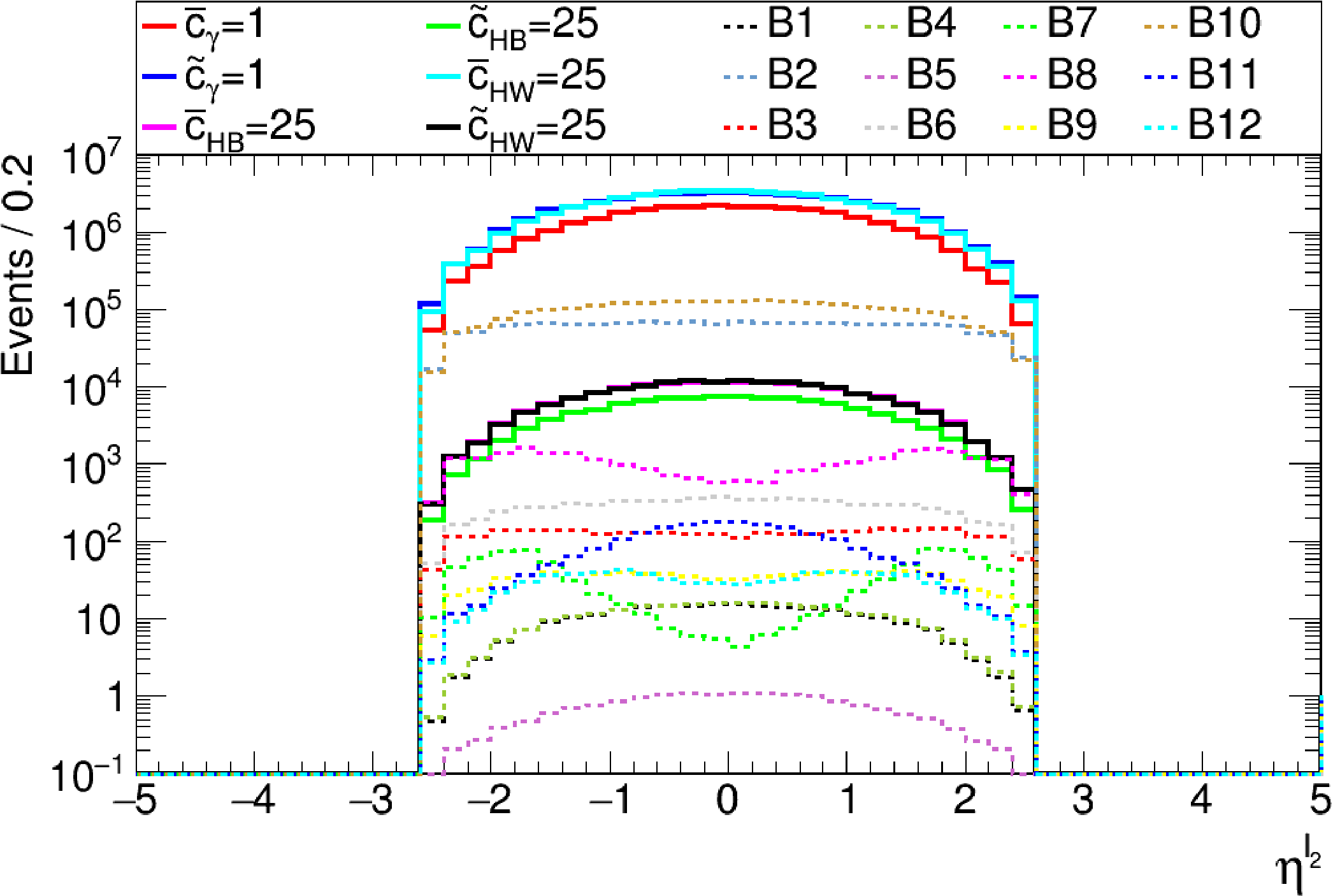}
\caption{}
\label{fig4:d}
\end{subfigure}\hfill

\caption{The pseudo-rapidity distributions of the leading charged lepton (left) and the sub-leading charged lepton (right) for signal with $\overline{c}_\gamma=1$, $\overline{c}_{HB}=25$, $\overline{c}_{HW}=25$, $\widetilde{c}_\gamma=1$, $\widetilde{c}_{HB}=25$, $\widetilde{c}_{HW}=25$ and relevant background
processes at the CLIC (top row) and the MuC (bottom row).}
\label{fig4}
\end{figure}

The leading and sub-leading charged leptons ($\ell_1$ and $\ell_2$) are ordered by their transverse momentum as $p_T^{\ell_1} > p_T^{\ell_2}$, respectively. The transverse momentum distributions of the leading and sub-leading charged leptons at the CLIC and the MuC for signal and relevant background processes are given in Fig.~\ref{fig3}. It can be seen from the distributions of both colliders that the signal can be separated from the backgrounds by $p_T^{\ell_1} > 80$ GeV and $p_T^{\ell_2} > 40$ GeV. Also, when the pseudo-rapidity distributions of the leading and sub-leading charged leptons for signal and relevant background processes in Fig.~\ref{fig4} are examined, deviations from the backgrounds begin to be seen around $|\eta^{\ell_{1,2}}|<2.4$ for CLIC and MuC. The coexistence of the transverse momentum and pseudo-rapidity distributions is labeled as Cut-1.

\begin{figure}[H]
\centering
\begin{subfigure}{0.48\linewidth}
\includegraphics[width=\linewidth]{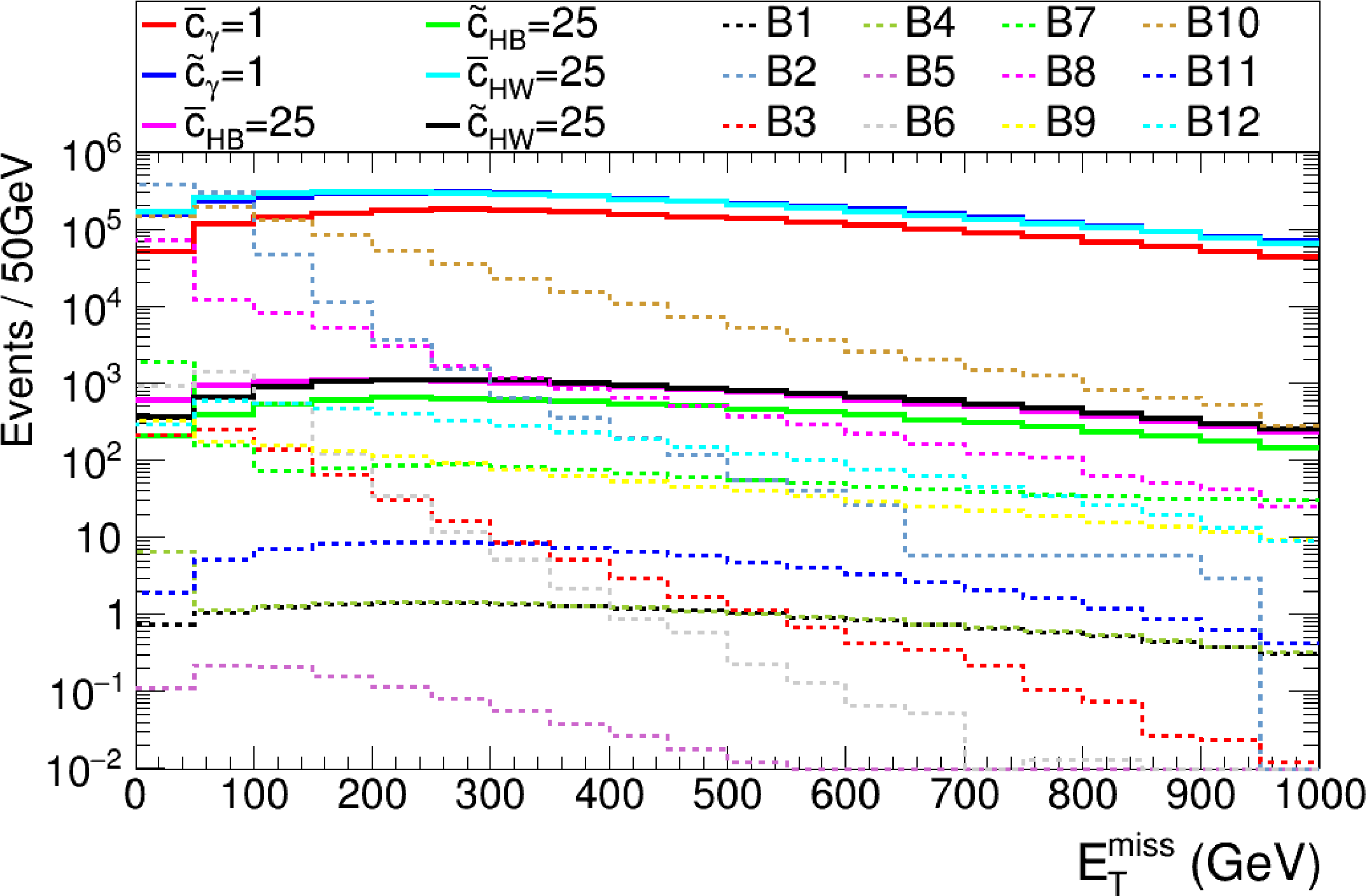}
\caption{}
\label{fig5:a}
\end{subfigure}\hfill
\begin{subfigure}{0.48\linewidth}
\includegraphics[width=\linewidth]{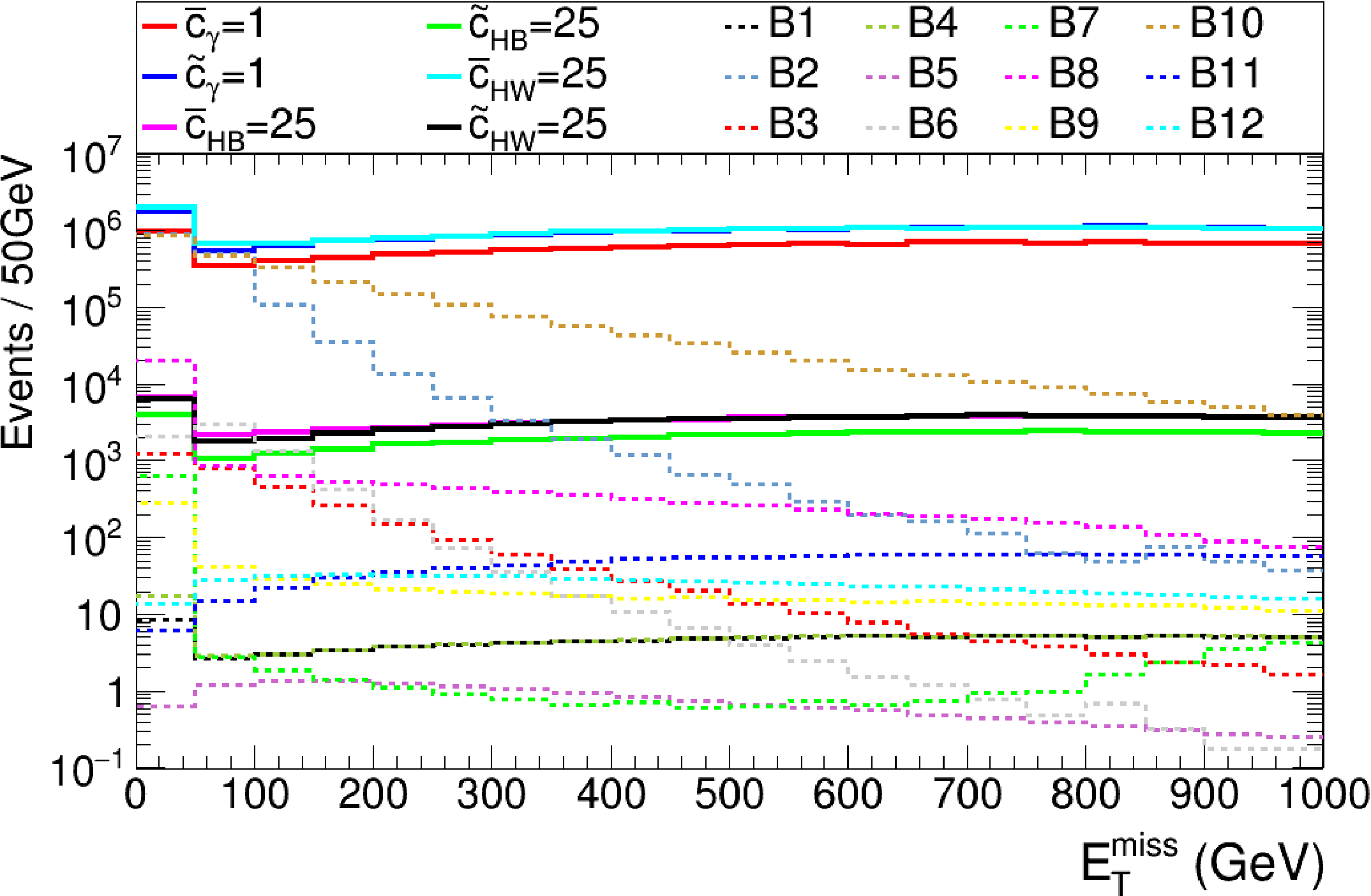}
\caption{}
\label{fig5:b}
\end{subfigure}\hfill

\caption{The distributions of missing transverse energy for signal with $\overline{c}_\gamma=1$, $\overline{c}_{HB}=25$, $\overline{c}_{HW}=25$, $\widetilde{c}_\gamma=1$, $\widetilde{c}_{HB}=25$, $\widetilde{c}_{HW}=25$ and relevant background
processes at the CLIC (left) and the MuC (right).}
\label{fig5}
\end{figure}

\begin{figure}[H]
\centering
\begin{subfigure}{0.48\linewidth}
\includegraphics[width=\linewidth]{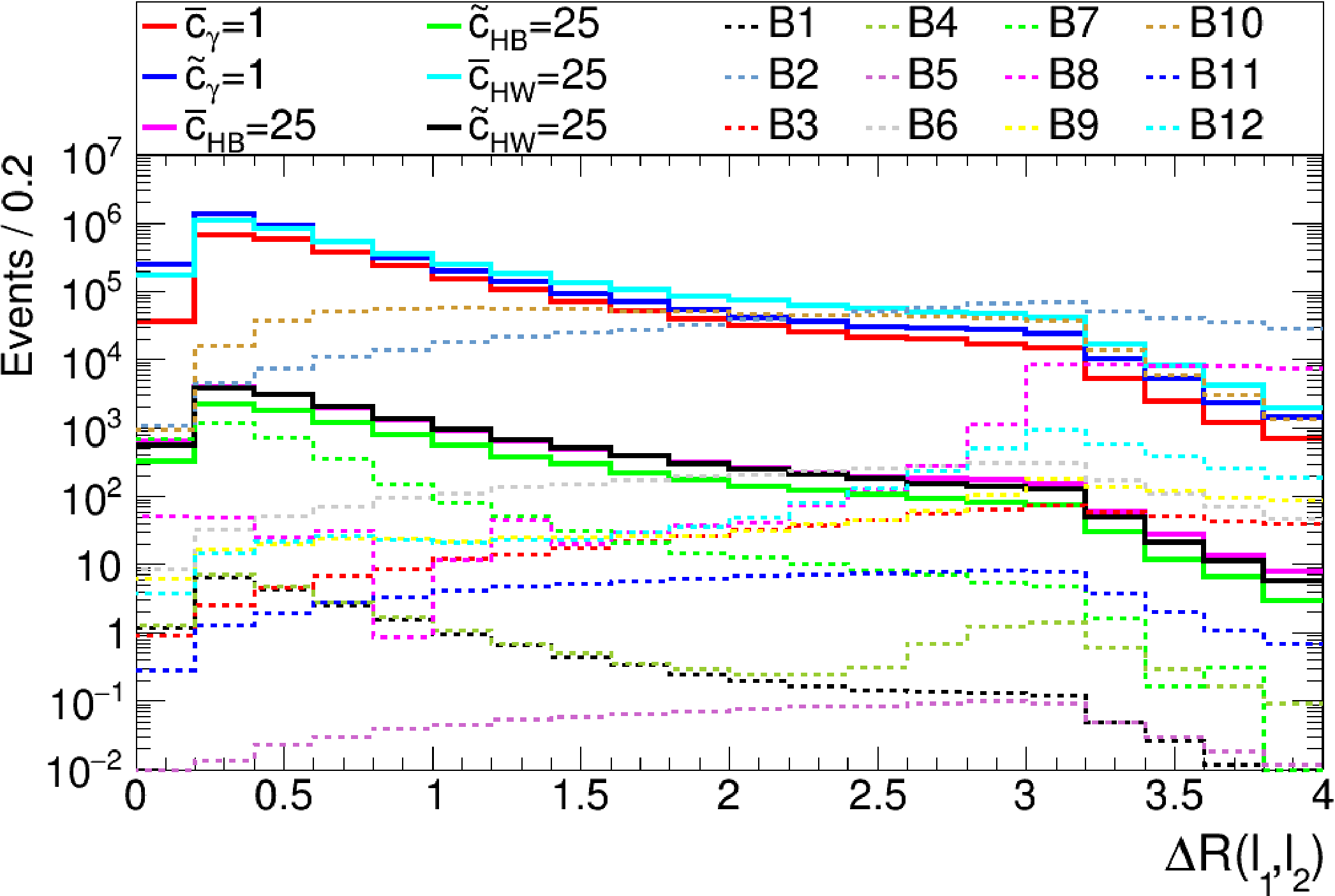}
\caption{}
\label{fig6:a}
\end{subfigure}\hfill
\begin{subfigure}{0.48\linewidth}
\includegraphics[width=\linewidth]{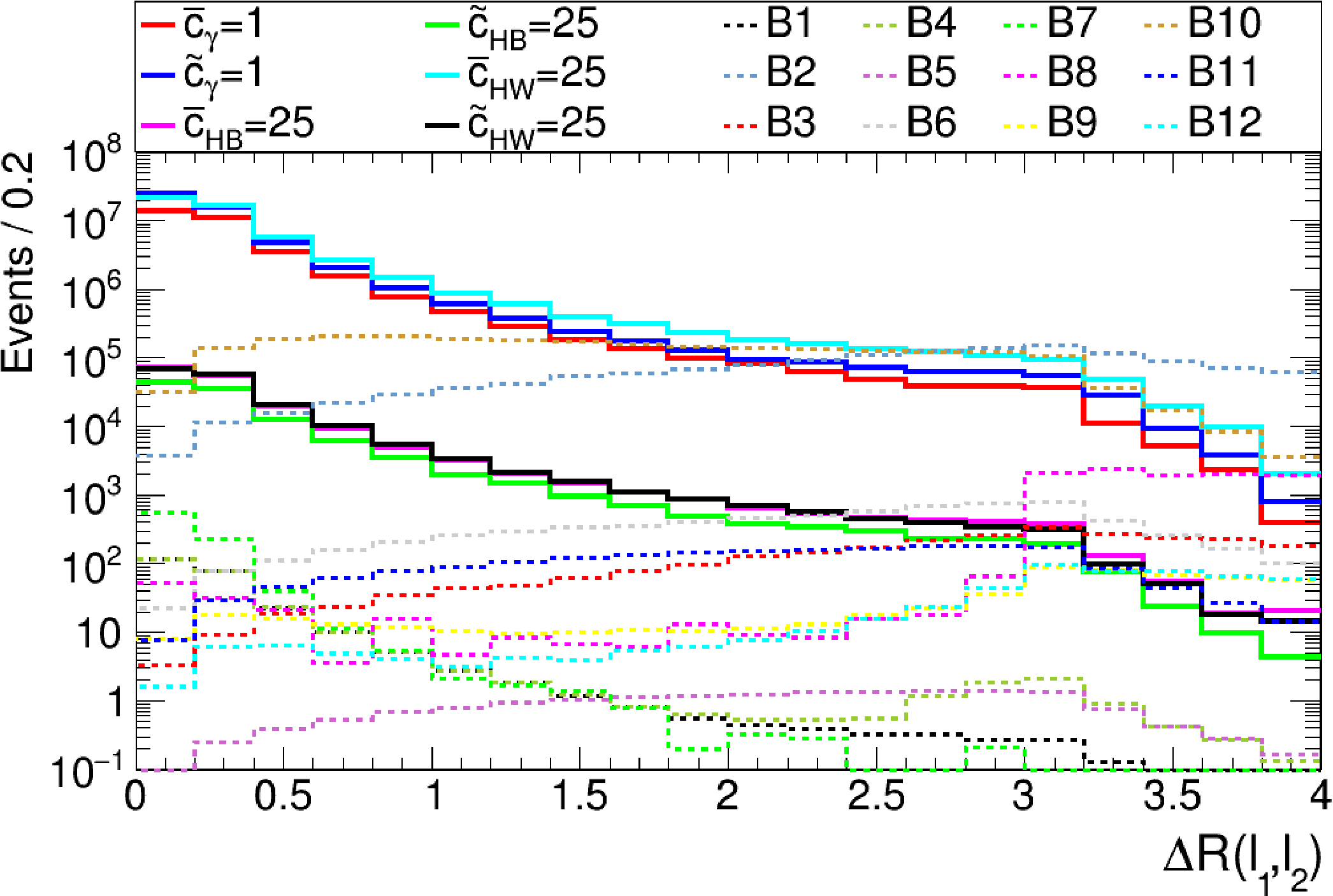}
\caption{}
\label{fig6:b}
\end{subfigure}\hfill

\caption{The distance between the leading and sub-leading charged leptons for signal with $\overline{c}_\gamma=1$, $\overline{c}_{HB}=25$, $\overline{c}_{HW}=25$, $\widetilde{c}_\gamma=1$, $\widetilde{c}_{HB}=25$, $\widetilde{c}_{HW}=25$ and relevant background processes at the CLIC (left) and the MuC (right).}
\label{fig6}
\end{figure}

\begin{figure}[H]
\centering
\begin{subfigure}{0.48\linewidth}
\includegraphics[width=\linewidth]{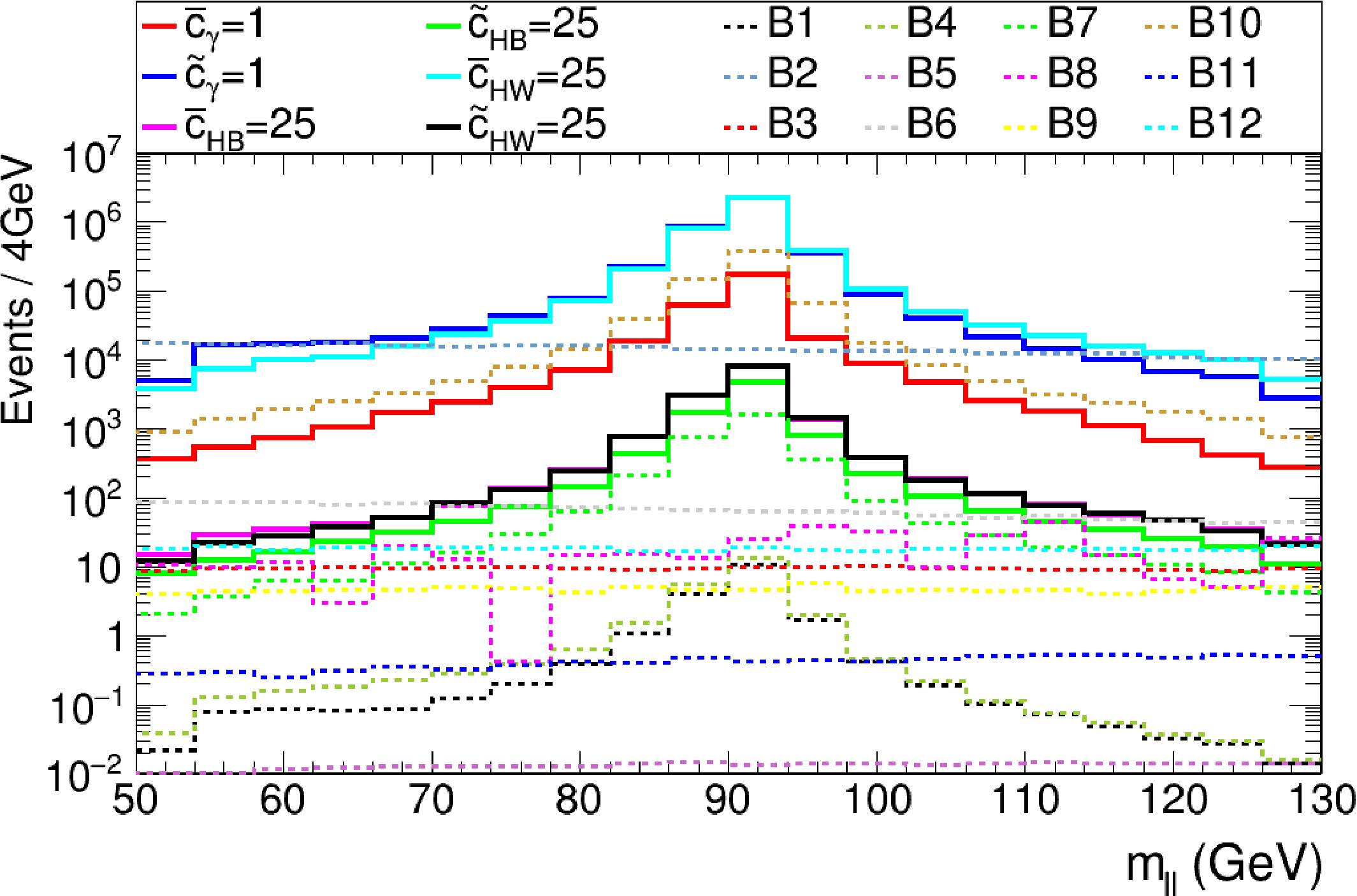}
\caption{}
\label{fig7:a}
\end{subfigure}\hfill
\begin{subfigure}{0.48\linewidth}
\includegraphics[width=\linewidth]{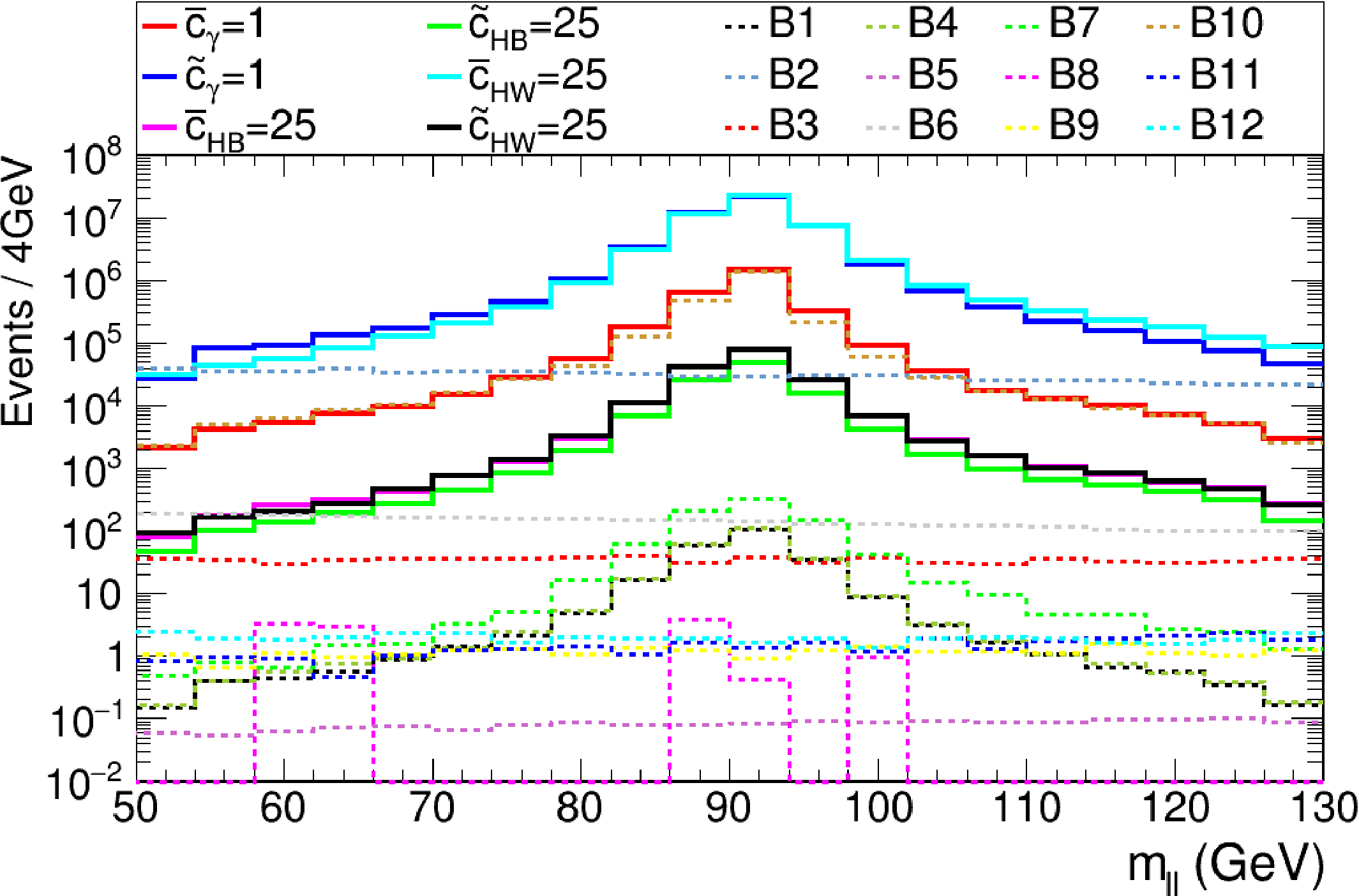}
\caption{}
\label{fig7:b}
\end{subfigure}\hfill

\caption{The invariant mass distributions of $Z$-boson decaying into the $\ell\ell$ channel for signal with $\overline{c}_\gamma=1$, $\overline{c}_{HB}=25$, $\overline{c}_{HW}=25$, $\widetilde{c}_\gamma=1$, $\widetilde{c}_{HB}=25$, $\widetilde{c}_{HW}=25$ and relevant background processes at the CLIC (left) and the MuC (right).}
\label{fig7}
\end{figure}

As can be seen from Figs.~\ref{fig5}-\ref{fig6}, a distinction between signals and backgrounds appears clearly at missing transverse energy $E_T^{miss} > 100$ GeV (Cut-2) and distance between the leading and sub-leading charged leptons $\Delta R({\ell_{1},\ell_{2}}) < 1.8$ (Cut-3). In order to distinguish the signal from the backgrounds in the invariant mass distribution of the dilepton pair resulting from the decay of the $Z$-boson in Fig.~\ref{fig7}, we determined the cuts to be in the range of $|m_{\ell\ell}-m_Z| < 20$ GeV for CLIC and MuC (Cut-4). When determining this cut, the $Z$-boson mass is considered to be $m_Z=91.2$ GeV.

In general, as can be seen in Figs.~\ref{fig3}-\ref{fig7}, the deviation regions between the signal and the background are similar for each distribution of CLIC and MuC. Therefore, the same kinematic cuts are determined in the analysis for CLIC and MuC and their definitions are summarized in Table~\ref{tab1}.

\begin{table}[H]
\centering
\caption{Summary of kinematic cuts used in the analysis of signal and background events at CLIC and MuC.}
\label{tab1}
\begin{tabular}{p{2.5cm}p{9.7cm}}
\hline \hline
Cuts & CLIC and MuC \\
\hline
Cut-0 & $N_{\ell\,(e,\,\mu)} >= 2$ at least two same flavor opposite-sign leptons\\ 
Cut-1 & $p_T^{\ell_1} > 80$ GeV, $p_T^{\ell_2} > 40$ GeV, $|\eta^{\ell_1}| < 2.4$, $|\eta^{\ell_2}| < 2.4$\\ 
Cut-2 & $E_T^{miss} > 100$ GeV\\
Cut-3 & $\Delta R({\ell_{1},\ell_{2}}) < 1.8$ \\
Cut-4 & $|m_{\ell\ell}-m_Z| < 20$ GeV\\ \hline \hline
\end{tabular}
\end{table}

The number of events after each cut used in the analysis for the signal ($\overline{c}_\gamma=1$, $\widetilde{c}_\gamma=1$, $\overline{c}_{HB}=25$, $\widetilde{c}_{HB}=25$, $\overline{c}_{HW}=25$,  $\widetilde{c}_{HW}=25$) and relevant SM backgrounds ($B1-B12$) at CLIC and MuC are presented in Table~\ref{tab2}. The number of events in this table are obtained by multiplying the cross-sections by the integrated luminosities, and the integrated luminosities are used as ${\cal L}_{\text{int}}=5$ ab$^{-1}$ for the CLIC and ${\cal L}_{\text{int}}=10$ ab$^{-1}$ for the MuC. To examine the effect of cuts on suppressing backgrounds, the ratio of the signal to the total background ($S/B_{tot}$) after each cut is given in Table~\ref{tab2}. It appears that the effect of Cut-3 and Cut-4 play a major role in suppressing the relevant backgrounds.

\begin{table}[H]
\centering
\caption{The number of events after applied kinematic cuts for signal and relevant background processes for CLIC and MuC.}
\label{tab2}
\begin{tabular}{p{2.1cm}p{1.4cm}p{1.2cm}p{1.4cm}p{1.2cm}p{1.4cm}p{1.2cm}p{1.4cm}p{1.2cm}p{1.4cm}p{1cm}}
\hline \hline
\multicolumn{11}{c}{CLIC} \\ [-0.5ex] \hline
Sig.\&Back. & Cut-0 & $S/B_{tot}$ & Cut-1 & $S/B_{tot}$ & Cut-2 & $S/B_{tot}$ & Cut-3 & $S/B_{tot}$ & Cut-4 & $S/B_{tot}$\\ [-0.5ex] \hline
$\overline{c}_{\gamma}=1$ & 152546 & 1.558 & 132775  & 4.579 & 128065  & 9.160 & 127658  & 11.07 & 28472.5  & 2.530 \\  [-1.5ex]
$\widetilde{c}_{\gamma}=1$ & 256196 & 2.617 & 215845 & 7.444 & 206782 & 14.79 & 206766 & 17.93 & 201907 & 17.94 \\  [-1.5ex]
$\overline{c}_{HB}=25$ & 928.556 & 0.009 & 672.068 & 0.023 & 641.487 & 0.045 & 641.381 & 0.055 & 625.031 & 0.055 \\  [-1.5ex]
$\widetilde{c}_{HB}=25$ & 537.466 & 0.005 & 372.815 & 0.012 & 360.515 & 0.025 & 360.476 & 0.031 & 351.571 & 0.031 \\  [-1.5ex] 
$\overline{c}_{HW}=25$ & 10032.5 & 0.102 & 7221.56 & 0.249 & 6887.62 & 0.492 & 6886.93 & 0.597 & 6713.31 & 0.596 \\  [-1.5ex]
$\widetilde{c}_{HW}=25$ & 5974.36 & 0.061 & 4138.88 & 0.142 & 3999.87 & 0.286 & 3999.36 & 0.346 & 3898.49 & 0.346 \\   [-0.5ex]
$B1$ & 1.22607 && 1.03185 && 0.98794 && 0.98787 && 0.96482 &\\ [-1.5ex]
$B2$ & 45541.7 && 8072.68 && 901.581 && 0.54663 && 0.36442 &\\ [-1.5ex]
$B3$ & 46.2573 && 16.5548 && 6.78068 && 0.35069 && 0.09216 &\\ [-1.5ex]
$B4$ & 1.60913 && 1.12543 && 0.97874 && 0.97866 && 0.95462 &\\ [-1.5ex]
$B5$ & 0.06597 && 0.04334 && 0.03175 && 0.01050 && 0.00196 &\\ [-1.5ex]
$B6$ & 192.431 && 38.7925 && 9.59501 && 0.25535 && 0.13781 &\\ [-1.5ex]
$B7$ & 211.458 && 168.913 && 66.8218 && 66.8218 && 65.1406 &\\ [-1.5ex]
$B8$ & 6646.37 && 6106.92 && 1329.27 && 0.74453 && 0 &\\ [-1.5ex]
$B9$ & 92.3525 && 74.9137 && 50.0290 && 5.50664 && 0.74897 &\\ [-1.5ex]
$B10$ & 44903.1 && 14326.6 && 11468.0 && 11452.6 && 11183.8 &\\ [-1.5ex]
$B11$ & 5.59499 && 4.88491 && 4.53798 && 1.53490 && 0.13245 &\\ [-1.5ex]
$B12$ & 250.385 && 181.844 && 141.562 && 1.17166 && 0.18441 &\\ 
\hline 

\multicolumn{11}{c}{MuC} \\ [-0.5ex] \hline
$\overline{c}_{\gamma}=1$ & 1398206 & 5.496 & 1326500 & 16.24 & 1277962 & 31.13 & 1276825 & 33.67 & 336033 & 9.104\\  [-1.5ex]
$\widetilde{c}_{\gamma}=1$ & 2538081 & 9.976 & 2389906 & 29.27 & 2294606 & 55.89 & 2294543 & 60.52 & 2243537 & 60.78 \\  [-1.5ex]
$\overline{c}_{HB}=25$ & 8224.37 & 0.032 & 7083.50 & 0.086 & 6772.93 & 0.164 & 6772.31 & 0.178 & 6611.50 & 0.179\\  [-1.5ex]
$\widetilde{c}_{HB}=25$ & 4947.52 & 0.019 & 4092.84 & 0.050 & 3925.45 & 0.095 & 3925.23 & 0.103 & 3829.65 & 0.103\\  [-1.5ex]
$\overline{c}_{HW}=25$ & 88588.7 & 0.348 & 76112.5 & 0.932 & 72711.2 & 1.771 & 72708.1 & 1.917 & 70956.2 & 1.922 \\  [-1.5ex]
$\widetilde{c}_{HW}=25$ & 55038.5 & 0.216 & 45382.1 & 0.555 & 43475.0 & 1.059 & 43472.8 & 1.146 & 42424.4 & 1.149 \\   [-0.5ex]
$B1$ & 12.1613 && 11.4756 && 11.0106 && 11.0102 && 10.7646 &\\  [-1.5ex]
$B2$ & 97890.6 && 19756.1 && 2576.04 && 0.78323 && 0 &\\  [-1.5ex]
$B3$ & 202.556 && 86.4668 && 33.6055 && 1.38968 && 0.29141 &\\  [-1.5ex]
$B4$ & 12.8208 && 11.7119 && 10.9933 && 10.9928 && 10.7606 &\\  [-1.5ex]
$B5$ & 1.03768 && 0.88521 && 0.79740 && 0.27493 && 0.02153 &\\  [-1.5ex]
$B6$ & 448.233 && 105.769 && 32.8345 && 0.88733 && 0.46603 &\\ [-1.5ex]
$B7$ & 25.6187 && 20.8526 && 5.08242 && 5.08242 && 4.97366 &\\  [-1.5ex]
$B8$ & 1577.13 && 1503.94 && 331.373 && 0.14782 && 0.05375 &\\  [-1.5ex]
$B9$ & 38.3228 && 34.5141 && 20.9155 && 1.79896 && 0.22648 &\\  [-1.5ex]
$B10$ & 154051 && 59989.1 && 37906.3 && 37853.1 && 36880.2 &\\  [-1.5ex]
$B11$ & 95.3493 && 92.4906 && 91.4581 && 24.9097 && 0.50677 &\\  [-1.5ex]
$B12$ & 43.2081 && 33.9411 && 31.9764 && 0.43080 && 0.02591 &\\
\hline \hline
\end{tabular}
\end{table}

\section{Sensitivities on the anomalous Higgs-gauge boson couplings} \label{Sec4}

The sensitivity of the dimension-six Higgs-gauge boson couplings in the $\gamma^* \gamma^* \rightarrow ZZ$ process is examined by applying the $\chi^2$ test \cite{Workman:2022ggb} to the simulated data. To obtain limits at the 95\% Confidence Level (C.L.), the $\chi^2$ distribution, where the $\chi^2$ critical values corresponding to degrees of freedom of 1 and 2 are equal to 3.84 and 5.99, respectively, is defined below:

\begin{eqnarray}
\label{eq.12}
\chi^{2}=\sum_{i}^{n_{bins}} (\frac{N_{i}^{TOT}-N_{i}^{B}}{N_{i}^{B}\Delta_{i}})^{2}
\end{eqnarray}

\begin{figure}[H]
\centering
\begin{subfigure}{0.48\linewidth}
\includegraphics[width=\linewidth]{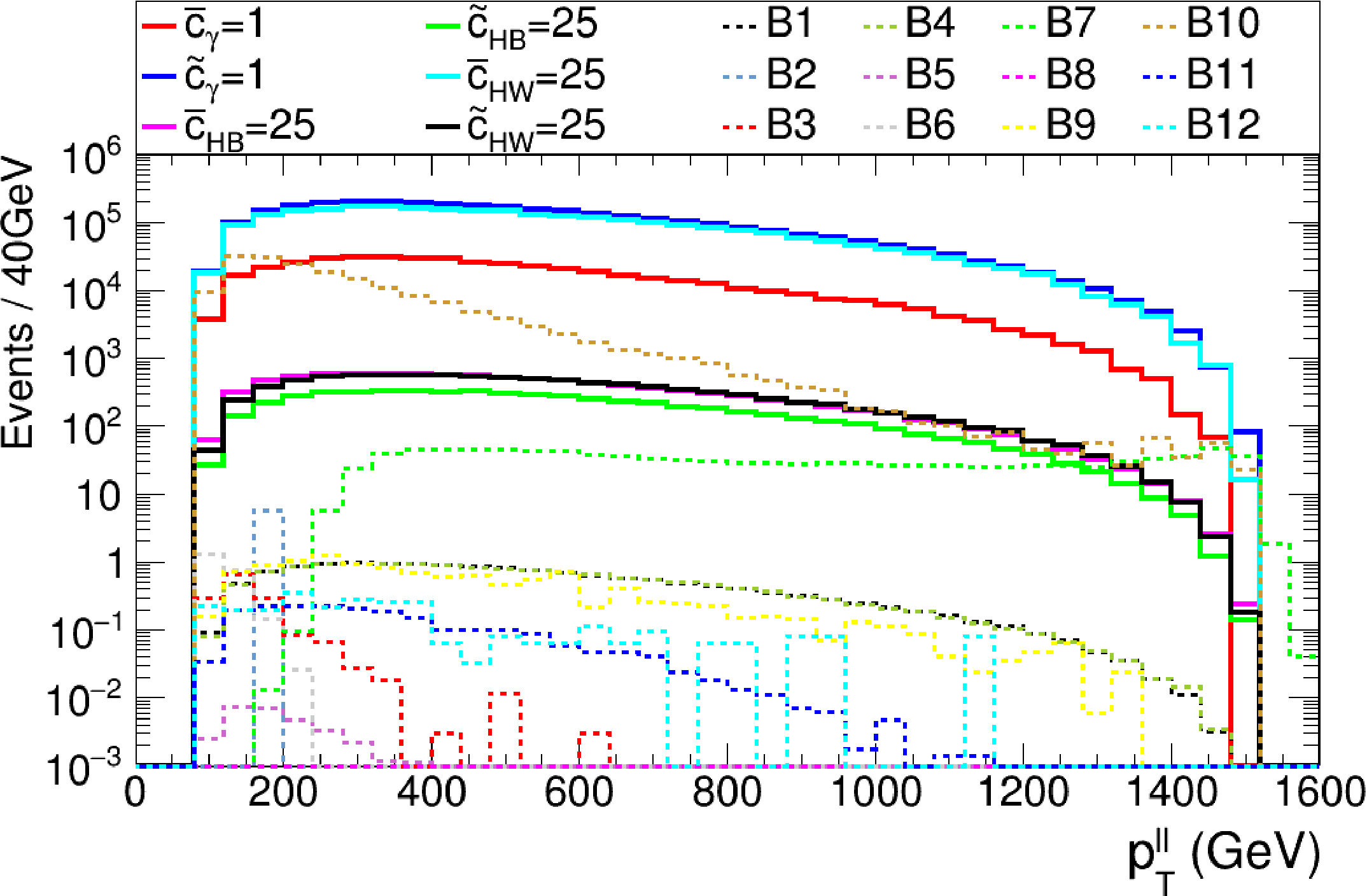}
\caption{}
\label{fig8:a}
\end{subfigure}\hfill
\begin{subfigure}{0.48\linewidth}
\includegraphics[width=\linewidth]{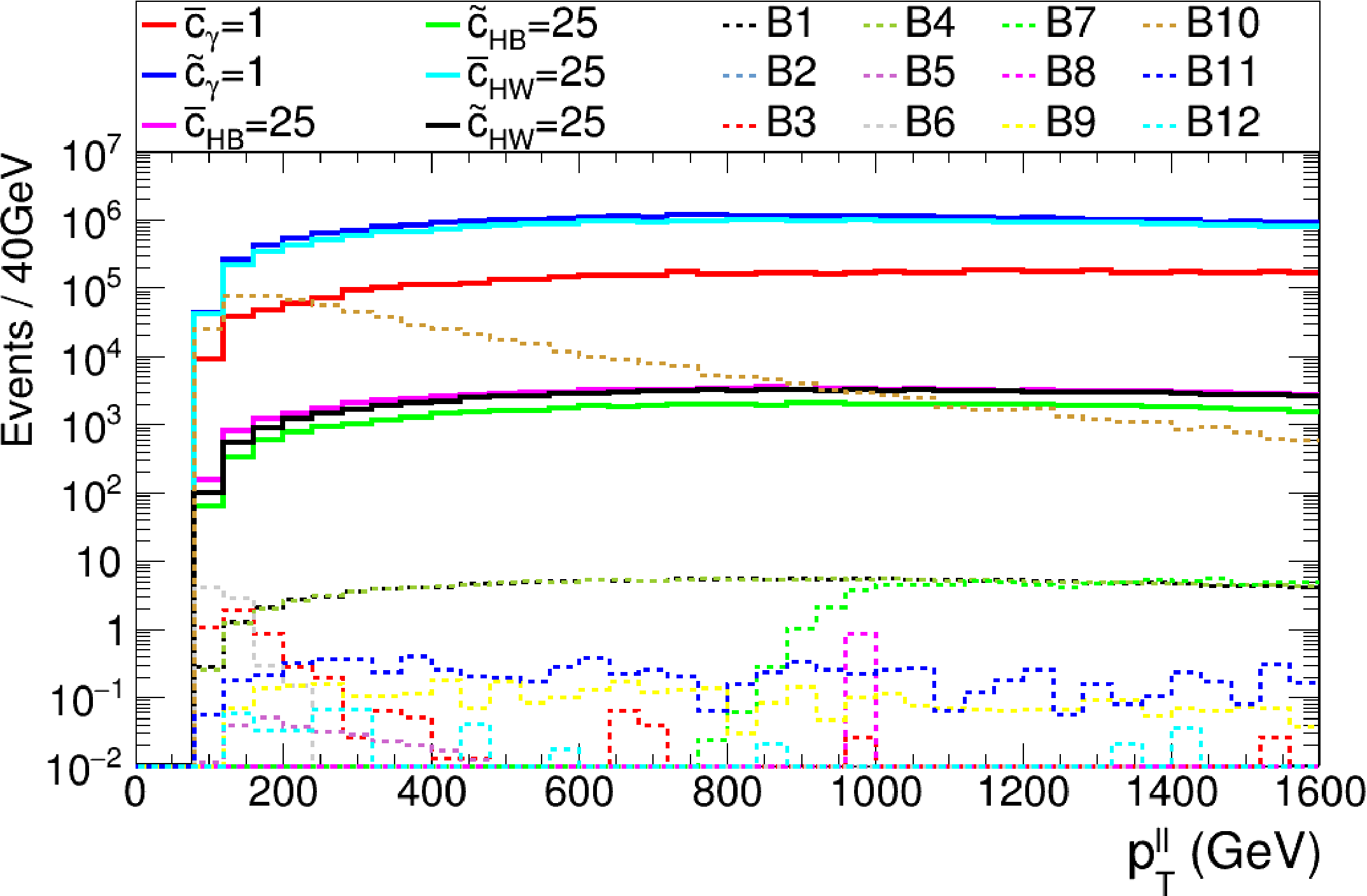}
\caption{}
\label{fig8:b}
\end{subfigure}\hfill

\caption{The transverse momentum distributions of dilepton $p_T^{\ell \ell}$ after applied cuts for signal with $\overline{c}_\gamma=1$, $\overline{c}_{HB}=25$, $\overline{c}_{HW}=25$, $\widetilde{c}_\gamma=1$, $\widetilde{c}_{HB}=25$, $\widetilde{c}_{HW}=25$ and relevant background processes at the CLIC (left) and the MuC (right).}
\label{fig8}
\end{figure}

{\raggedright where $N_{i}^{TOT}$ is the total number of events including contributions of effective couplings and $N_{i}^{B}$ is the number of events of relevant backgrounds in ith bin of the dilepton transverse momentum $p_T^{\ell \ell}$ distribution. $\Delta_{i}=\sqrt{\delta_{sys}^2+\frac{1}{N_{i}^{B}}}$ is the sum of systematic and statistical errors in each bin. The transverse momentum distributions of dilepton $p_T^{\ell \ell}$ after applied cuts for signal and relevant background processes at CLIC and MuC are given in Fig.~\ref{fig8}.} 

The 95\% C.L. limits of $\overline{c}_\gamma$, $\widetilde{c}_\gamma$, $\overline{c}_{HB}$, $\widetilde{c}_{HB}$, $\overline{c}_{HW}$ and $\widetilde{c}_{HW}$ coefficients with and without systematic error of 5\% for 3 and 10 TeV center-of-mass energies with integrated luminosities of 5 ab$^{-1}$ and 10 ab$^{-1}$ at CLIC and MuC, respectively, are given in Table~\ref{tab3}. It is seen that the most sensitive limits among all the coefficients are obtained with MuC.

\begin{table}[H]
\centering
\caption{The 95\% C.L. limits on the anomalous Higgs-gauge boson couplings at CLIC, MuC, other experimental and phenomenological studies.}
\label{tab3}
\centering
\begin{tabular}{p{2.4cm}p{1.1cm}p{3.1cm}p{3.5cm}p{1.9cm}p{3.2cm}}
\hline \hline
Coefficients & $\delta_{sys}$ & CLIC & MuC & \multicolumn{2}{l}{Other Studies and Limits} \\ \hline
\multirow{6}{*}{$\overline{c}_{\gamma}$}  
& 0\% & [-0.0096; 0.0141] & [-0.0023; 0.0065] & Ref. \cite{Aad:2016hws} & [-0.00074; 0.00057]\\
& 5\% & [-0.0143; 0.0188] & [-0.0049; 0.0093] & Ref. \cite{ATLAS:2019sdf} & [-0.00011; 0.00011]\\
& & & & Ref. \cite{Englert:2016onz} & [-0.00016; 0.00013]\\
& & & & Ref. \cite{Khanpour:2017ubw} & [-0.01909; 0.00625]\\
& & & & Ref. \cite{Denizli:2019oxc} & [-0.0051; 0.0038]\\
& & & & Ref. \cite{Denizli:2021uhb} & [-0.0089; 0.0066]\\ \hline
\multirow{4}{*}{$\widetilde{c}_{\gamma}$} 
& 0\% & [-0.0114; 0.0114] & [-0.0039; 0.0039] & Ref. \cite{Aad:2016hws} & [-0.0018; 0.0018]\\
& 5\% & [-0.0163; 0.0162] & [-0.0067; 0.0067] & Ref. \cite{ATLAS:2019sdf} & [-0.00028; 0.00043]\\
& & & & Ref. \cite{Denizli:2019oxc} & [-0.0043; 0.0043]\\
& & & & Ref. \cite{Denizli:2021uhb} & [-0.0077; 0.0077]\\ \hline
 \multirow{2}{*}{$\overline{c}_{HB}$}
& 0\% & [-5.9449; 4.7223] & [-2.5608; 1.3256] & Ref. \cite{Englert:2016onz} & [-0.004; 0.004]\\
& 5\% & [-8.0895; 6.8177] & [-3.8492; 2.5888] & Ref. \cite{Khanpour:2017ubw} & [-0.0172; 0.00661]\\ \hline
\multirow{2}{*}{$\widetilde{c}_{HB}$} 
& 0\% & [-6.5370; 6.5680] & [-2.3449; 2.3621] & & \\ 
& 5\% & [-9.2347; 9.2712] & [-3.9892; 3.9971] & & \\ \hline
\multirow{5}{*}{$\overline{c}_{HW}$} 
& 0\% & [-1.7723; 1.4080] & [-0.7692; 0.3969] & Ref. \cite{Aad:2016hws} & [-0.086; 0.092]\\
& 5\% & [-2.3400; 2.0478] & [-1.1405; 0.7739] & Ref. \cite{Aaboud:2018yer} & [-0.057; 0.051]\\
& & & & Ref. \cite{ATLAS:2019sdf} & [-0.025; 0.022]\\
& & & & Ref. \cite{Englert:2016onz} & [-0.004; 0.004]\\
& & & & Ref. \cite{Khanpour:2017ubw} & [-0.00187; 0.0018]\\ \hline 
\multirow{3}{*}{$\widetilde{c}_{HW}$} 
& 0\% & [-2.0053; 1.995] & [-0.6968; 0.6964] & Ref. \cite{Aad:2016hws} & [-0.23; 0.23]\\
& 5\% & [-2.7855; 2.7752] & [-1.1946; 1.1887] & Ref. \cite{Aaboud:2018yer} & [-0.16; 0.16]\\ 
& & & & Ref. \cite{ATLAS:2019sdf} & [-0.065; 0.063]\\ 
   \hline \hline
\end{tabular}
\end{table}

\begin{figure}[H]
\centering
\begin{subfigure}{0.51\linewidth}
\includegraphics[width=\linewidth]{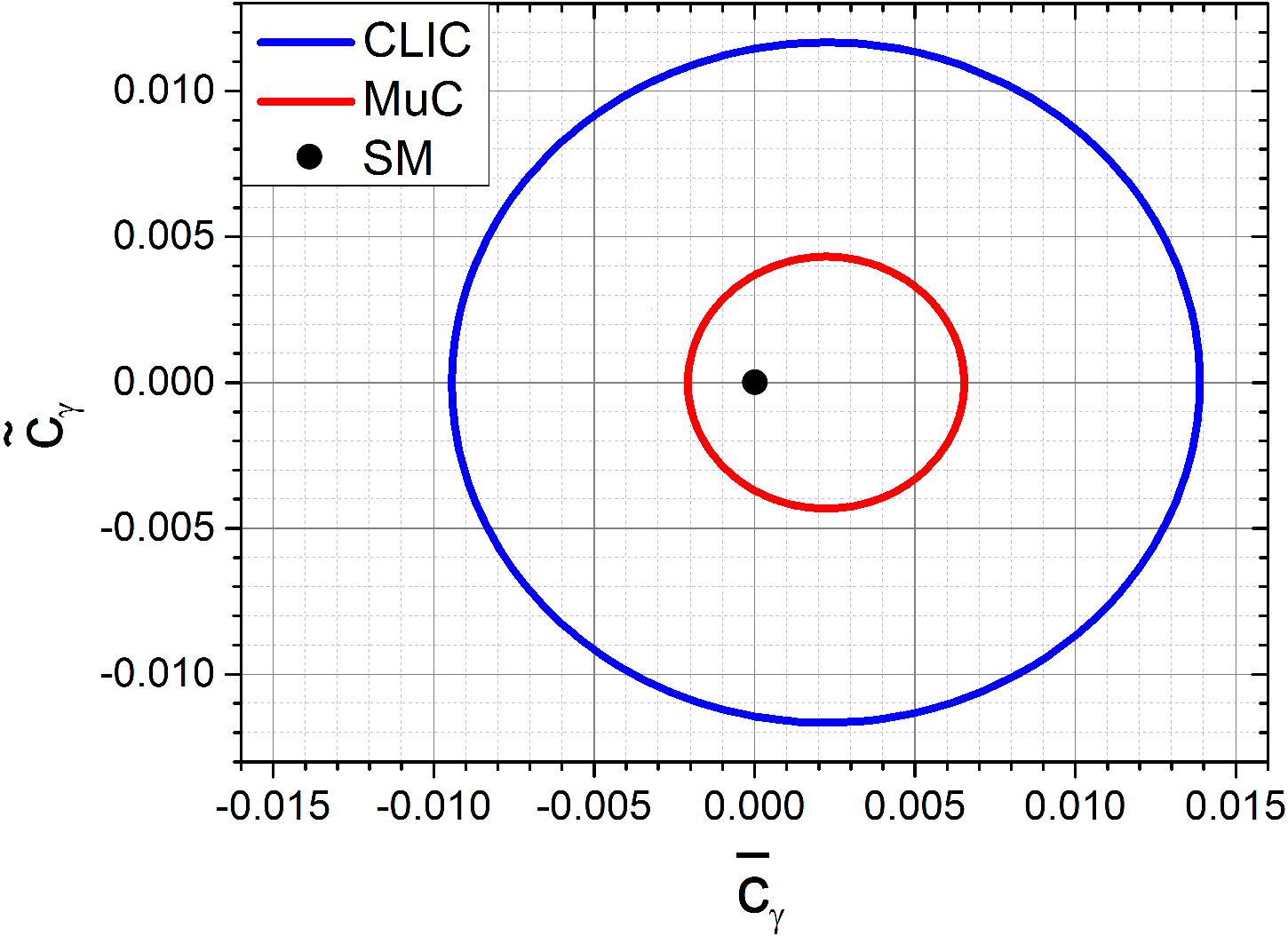}
\caption{}
\label{fig9:a}
\end{subfigure}\hfill
\begin{subfigure}{0.48\linewidth}
\includegraphics[width=\linewidth]{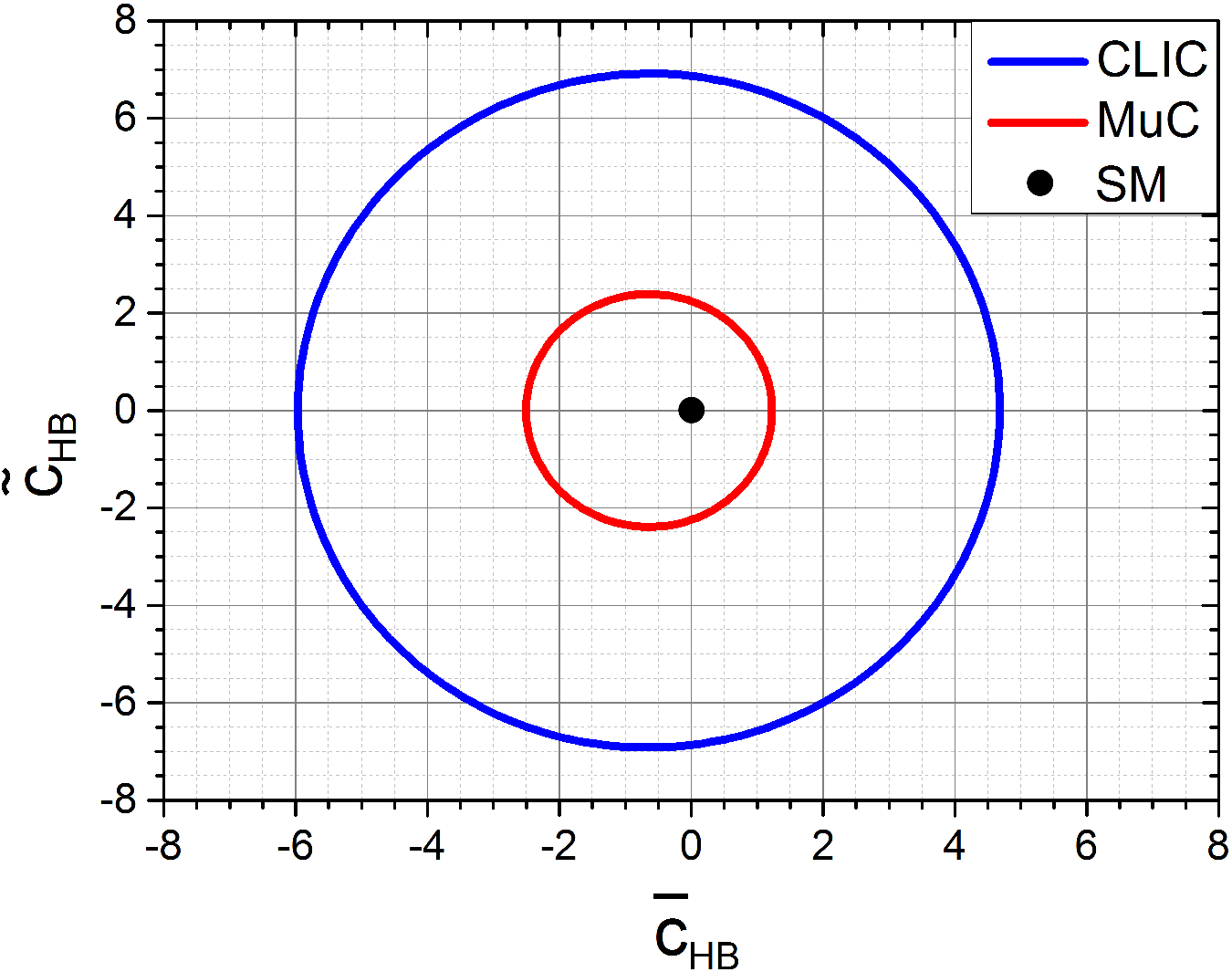}
\caption{}
\label{fig9:b}
\end{subfigure}\hfill
\begin{subfigure}{0.48\linewidth}
\includegraphics[width=\linewidth]{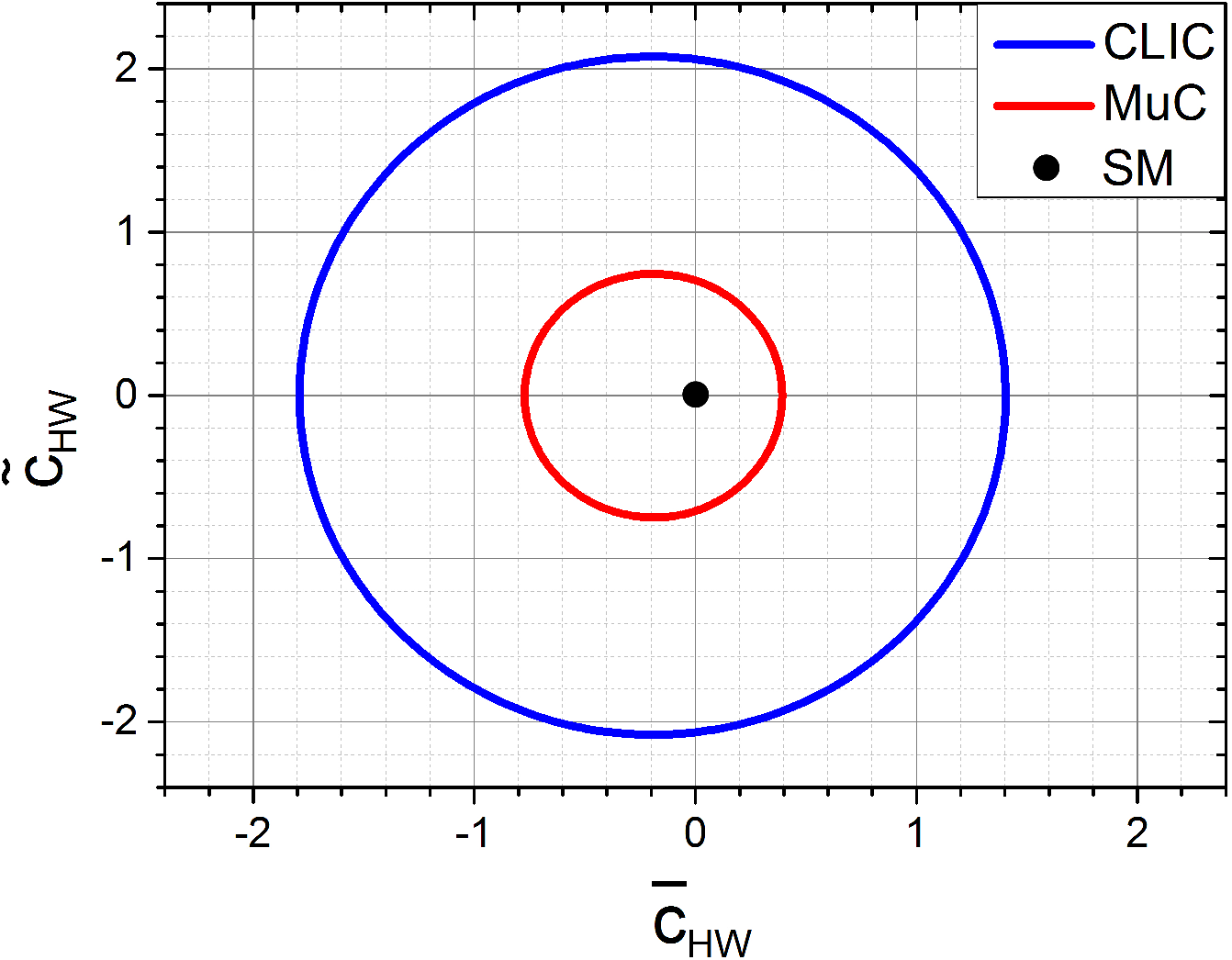}
\caption{}
\label{fig9:c}
\end{subfigure}\hfill

\caption{Two-dimensional 95\% C.L. intervals in plane for $\overline{c}_\gamma-\widetilde{c}_\gamma$ (a), $\overline{c}_{HB}-\widetilde{c}_{HB}$ (b) and $\overline{c}_{HW}-\widetilde{c}_{HW}$ (c) with taking systematic error $\delta_{sys}=0\%$ at the CLIC and the MuC. The black point markers represent the SM expectation.}
\label{fig9}
\end{figure}

The one-parameter analysis discussed above is useful in scenarios where five of the six non-trivial coefficients can be severely constrained. To understand how the constraints on a particular CP-violating coupling change in the presence of CP-conserving couplings, which is also the focus of this study, we consider the case in which two of the six Wilson coefficients are considered to be non-zero. Fig.~\ref{fig9} shows the 95\% C.L. contours in the $\overline{c}_\gamma-\widetilde{c}_\gamma$, $\overline{c}_{HB}-\widetilde{c}_{HB}$ and $\overline{c}_{HW}-\widetilde{c}_{HW}$  planes at CLIC and MuC with the two-parameter analysis.

\section{Conclusions} \label{Sec5}

The future lepton colliders CLIC and MuC are attractive options with a clean environment for a better understanding of Higgs interactions in new physics beyond the SM. We have investigated the CP-conserving and CP-violating dimension-six operators of Higgs-gauge boson couplings through the process $\gamma^* \gamma^* \rightarrow ZZ$ using an effective Lagrangian approach at CLIC with $\sqrt{s}=3$ TeV and ${\cal L}_{\text{int}}=5$ ab$^{-1}$ and MuC with $\sqrt{s}=10$ TeV and ${\cal L}_{\text{int}}=10$ ab$^{-1}$. We have analyzed the $\ell\ell\nu\nu$ decay channel of the $Z$-boson pair considering realistic detector effects with tuned CLIC and MuC detector cards in the analysis. We have shown the kinematic distributions of the signal and relevant background processes to determine a cut-based analysis: transverse momentum, pseudo-rapidity, invariant mass and distance of the leading and sub-leading leptons, and missing transverse energy of two neutrinos. The reason because it aims to find the optimum cuts to obtain the best limits on the Wilson coefficients for the anomalous $H\gamma\gamma$ and $HZZ$ couplings. We have obtained 95\% C.L. limits on the $\overline{c}_\gamma$, $\widetilde{c}_\gamma$, $\overline{c}_{HB}$, $\widetilde{c}_{HB}$, $\overline{c}_{HW}$ and $\widetilde{c}_{HW}$ Wilson coefficients of the SILH basis by analyzing the distributions of dilepton transverse momentum in the signal and relevant background processes.

The 95\% C.L. limits on Wilson coefficients in the $H\rightarrow \gamma\gamma$ decay channel were investigated by the ATLAS experiment with an integrated luminosity of 20.3 fb$^{-1}$ at $\sqrt{s}=8$ TeV, and the experimental limits of $\overline{c}_\gamma$, $\widetilde{c}_\gamma$, $\overline{c}_{HW}$ and $\widetilde{c}_{HW}$ coefficients were obtained as $[-0.00074; 0.00057] \cup [0.0038; 0.0051]$, $[-0.0018; 0.0018]$, $[-0.086; 0.092]$ and $[-0.23; 0.23]$, respectively \cite{Aad:2016hws}. In a similar analysis at $\sqrt{s}=13$ TeV with an integrated luminosity of 36.1 fb$^{-1}$, ATLAS collaboration stated that the $H\rightarrow \gamma\gamma$ decay channel is not sensitive to $\overline{c}_\gamma$ and $\widetilde{c}_\gamma$, and determined the experimental limits of $\overline{c}_{HW}$ and $\widetilde{c}_{HW}$ coefficients to $[-0.057; 0.051]$ and $[-0.16; 0.16]$, respectively \cite{Aaboud:2018yer}. In the study where the integrated luminosity was increased to 139 fb$^{-1}$ at $\sqrt{s}=13$ TeV, the limits for $\overline{c}_\gamma$, $\widetilde{c}_\gamma$, $\overline{c}_{HW}$ and $\widetilde{c}_{HW}$ coefficients were obtained as $[-0.00011; 0.00011]$, $[-0.00028; 0.00043]$, $[-0.025; 0.022]$ and $[-0.065; 0.063]$, respectively, by the ATLAS collaboration \cite{ATLAS:2019sdf}. 

In some phenomenological studies, limits are given at 95\% C.L. for the 14 TeV LHC with 3000 fb$^{-1}$ in Ref.~\cite{Englert:2016onz} and for the 350 GeV electron positron collider with 3000 fb$^{-1}$ in Ref.~\cite{Khanpour:2017ubw}. The limit of $\overline{c}_\gamma$ coefficient in CLIC and MuC is 1.1 and 2.8 times more sensitive, respectively, than in Ref.~\cite{Khanpour:2017ubw}, but the other limits are worse. The limits at 95\% C.L. on $\overline{c}_\gamma$ and $\widetilde{c}_\gamma$ coefficients through the process $pp\rightarrow\gamma\gamma\gamma$ were reported as $[-0.0051; 0.0038]$ and $[-0.0043; 0.0043]$ at FCC-hh with 100 TeV/10 ab$^{-1}$ \cite{Denizli:2019oxc} and $[-0.0089; 0.0066]$ and $[-0.0077; 0.0077]$ at LE-FCC with 37.5 TeV/15 ab$^{-1}$ \cite{Denizli:2021uhb}. Compared to Ref.~\cite{Denizli:2019oxc,Denizli:2021uhb}, our limit on $\overline{c}_\gamma$ and $\widetilde{c}_\gamma$ coefficients at MuC is more sensitive. The limits of all the mentioned phenomenological and experimental studies on anomalous Higgs-gauge boson couplings are summarized in Table~\ref{tab3}. As a result of the findings of this study, future CLIC and MuC with photon-induced interactions may improve the sensitivity limits on $\overline{c}_\gamma$, $\widetilde{c}_\gamma$, $\overline{c}_{HB}$, $\widetilde{c}_{HB}$, $\overline{c}_{HW}$ and $\widetilde{c}_{HW}$ coefficients for anomalous $H\gamma\gamma$ and $HZZ$ couplings relative to the experimental limits of the LHC and the phenomenological limits of future hadron-hadron colliders.

\end{document}